\documentclass[12pt,preprint]{aastex}

\newcommand{\iso}{{\em ISO}}

\newcommand{\mum}{\ifmmode{\rm \mu m}\else{$\mu$m}\fi}
\newcommand{\er}{\ifmmode{\pm}\else{$\pm$}\fi}

\usepackage{graphicx}

\slugcomment{Accepted for publication in the Astrophysical Journal, 2008}
\shorttitle{PAHs in Herbig AeBe Stars}
\shortauthors{Keller et al.}

\begin{document}

\title{PAH emission from Herbig AeBe stars}

\author{Luke~D.~Keller\altaffilmark{1}, G.~C.~Sloan\altaffilmark{2}, W.~J.~Forrest\altaffilmark{3},  S.~Ayala\altaffilmark{4}, P.~D'Alessio\altaffilmark{4}, S.~Shah\altaffilmark{1}, N.~Calvet\altaffilmark{5},  J.~Najita\altaffilmark{6},
A.~Li\altaffilmark{7}, L.~Hartmann\altaffilmark{5}, B.~Sargent\altaffilmark{3}, D.~M.~Watson\altaffilmark{3}, \& C.~H.~Chen\altaffilmark{8}}


\altaffiltext{1}{Department of Physics, Ithaca College, Ithaca,
NY 14850} 
\altaffiltext{2}{Cornell University, Astronomy Department,
Ithaca, NY 14853-6801} 
\altaffiltext{3}{Department
of Physics and Astronomy, University of Rochester, Rochester, NY 14627-0171}
\altaffiltext{4}{Centro de Radioastronomia y Astrofisica, UNAM, 
Apartado Postal 3-72 (Xangari), 58089 Morelia, Michoacan, Mexico}
\altaffiltext{5}{Department
of Astronomy, University of Michigan, Ann Arbor, MI 48109}
\altaffiltext{6}{National
Optical Astronomy Observatory, 950 North Cherry Avenue, Tucson, AZ
85719}
\altaffiltext{7}{Department of Physics \& Astronomy, University
of Missouri, Columbia, MO 65211} 
\altaffiltext{8}{Space Telescope Science Institute, 3700 San Martin Drive
Baltimore, MD 21218}


\begin{abstract}
We present spectra of a sample of Herbig Ae and Be (HAeBe) stars obtained
with the Infrared Spectrograph on the \textit{Spitzer Space Telescope}. All but one of the Herbig
stars show emission from polycyclic aromatic hydrocarbons (PAHs) and seven of
the spectra show PAH emission, but no silicate emission at 10~\mum. The central wavelengths of the 6.2, 7.7--8.2, and 11.3~\mum\ emission features decrease with stellar temperature,
indicating that the PAHs are less photo-processed in cooler radiation fields.
The apparent low level of photo processing in HAeBe stars, relative to other PAH emission sources, implies that the PAHs are newly exposed to the UV-optical radiation fields from
their host stars. HAeBe stars show a variety of PAH emission intensities and ionization fractions, but a narrow range of PAH spectral classifications based on positions of major
PAH feature centers. This may indicate that, regardless of their locations relative
to the stars, the PAH molecules are altered by the same physical processes in the proto-planetary disks of intermediate-mass stars. Analysis of the mid-IR spectral energy distributions indicates that our sample likely includes both radially flared and more flattened/settled disk systems, but we do not see the expected correlation of overall PAH emission with disk geometry.
We suggest that the strength of PAH emission from HAeBe stars may depend not only on the degree of radial flaring, but also on the abundance of PAHs in illuminated regions of the disks and possibly on the vertical structure of the inner disk as well.
\end{abstract}


\section{INTRODUCTION}

Circumstellar accretion disks that contain dust and gas are common
around intermediate mass stars. Studying the material in these disks
and their evolution can aid us in understanding the very latest stages
of star formation and the early stages of planet formation. Herbig Ae/Be 
(HAeBe) stars are intermediate-mass (M$\sim$2-10 M$_{\sun}$) 
pre-main-sequence stars of spectral class A, B, or early F 
\citep{her60, str72, the94, mal98}. Though often associated with nebulosity,
HAeBe photospheres are directly observable at visible wavelengths.
Unlike normal A and B stars, HAeBe stars have strong emission
lines (e.g. H$\alpha$ and Br$\gamma$) in their spectra, they can be highly variable
at visible wavelengths, and their infrared fluxes are strongly in
excess of purely photospheric emission. Current theory predicts that
intermediate-mass stars should form more quickly than lower-mass (T
Tauri) stars, but they will not reach their stellar birth lines heavily
enshrouded in dense circumstellar envelopes as high-mass stars do.
This means that we can directly view the stellar photospheres and
residual accretion disks of HAeBe stars, with a relatively unobscured
view of both. Unlike high-mass stars, while some HAeBe stars are associated
with star forming regions they are not common in the extreme environments
of HII regions near OB associations. Thus the stars themselves, and
not their environments, drive their disk structure, chemistry, and evolution.

Chemical and dynamical studies of Galactic HAeBe stars have shown
holes and gaps in their inner disks indicating the possible dynamical influence of newly
formed planets \citep[e.g.][]{gra05,bri07}, as well as evidence of different disk geometries
from those that flare
with increasing radius and have dust with a large abundance of small
grains, to flattened disks composed of larger dust grains \citep[e.g.][]{mee01,aa04}. These
characteristics are similar to those of the lower-mass T Tauri systems. 

Solid-state features and thermal radiation from warm dust, as well
as emission from atomic and molecular gas all produce
measurable signatures in the mid-IR spectra of young stars with
circumstellar disks. Mid-IR imaging and spectroscopy, much of it
from space telescopes, have
therefore played a major role in measuring and characterizing 
the disks orbiting T Tauri and HAeBe stars. Unlike T
Tauri stars, of which only $\sim$8\% show emission from
polycyclic aromatic hydrocarbons
\citep[PAHs,][]{gee06}, nearly 50\% of HAeBe stars have strong PAH emission
\citep[e.g.][]{mee01,aa04}, which requires a direct line of sight
from the emitting material---presumably located on the surfaces and/or
inner rims of disks---to the photospheric optical and UV radiation
fields. Furthermore, spectra from the Infrared Space Observatory (\iso)
and recent \textit{Spitzer} observations have revealed that many HAeBe
stars in the Galaxy have large abundances of crystalline silicate
grains in their disks indicating significant thermal dust processing
subsequent to the formation of the star-disk system. These latter
characteristics are not as common in T Tauri disks. Moreover, an evolutionary
link from HAeBe proto-planetary disks to Vega-like debris disks is
becoming evident in studies of large samples of HAeBe stars 
\citep[e.g.][]{her06,gra07}, but the details have yet
to be completely worked out. Though details of time scale and evolution
remain elusive, recent modeling efforts have succeeded
in predicting spectral energy distributions from varying physical characteristics
in the disks like dust grain size and composition, properties of the stellar radiation field, 
disk heating and cooling, and disk geometry \citep[e.g.][]{dul04,dal06,dul07,rob07}.

It is remarkable that such a large fraction of HAeBe stars are
PAH emission sources. \cite{mee01} suggested that PAH emission
may change measurably with the structure and physical properties of
the disks and perhaps with disk evolution. PAHs have been observed in many different
astrophysical contexts: the Galactic and extra-galactic ISM, dust envelopes 
surrounding post-AGB (asymptotic giant branch) stars, planetary nebulae (PNe),
and young stellar objects (YSOs). The molecular structure of PAHs can differ significantly
depending on the physical characteristics of the objects or environments
where they are found. The bottom line is that PAHs are abundant in
interstellar and circumstellar environments, and they can be important
constituents in the energy balance of those environments as sources of 
photoelectrons that heat the gas component \citep{kam04}. 

Emission from PAH molecules may
also trace proto-planetary disk geometry \citep{mee01,aa04,gra05}. Exactly what the
PAH emission can tell us about the disk environments will depend upon
a thorough and detailed understanding of how the molecules are energized and
processed by the stellar radiation fields and how they
are distributed radially and vertically in the disks. Throughout this paper when
we refer to "processing" of PAHs, we mean photo-processing by which 
the optical and UV stellar radiation break molecular bonds producing smaller
PAH molecules or completely destroying them.
We have begun a detailed analysis of PAH emission from HAeBe stars
using very high S/N spectra that allow much higher precision in analyzing
PAH feature strengths and wavelengths, and thus the molecular structure
that produces them, than has been possible before.

In a previous paper (Sloan et. al. 2005, hereafter Paper 1), we analyzed
four HAeBe stars in the 5-14~\mum\ range that have strong PAH emission
but no silicate peak at 10~\mum. We identified a trend towards decreasing
PAH molecule size with increasing PAH ionization fraction and noted
that the center wavelengths of the PAH features shift systematically
with increasing UV field strength from the stars. We suggested that
HAeBe stars may have distinctive PAH spectra relative to other
PAH sources (e.g. interstellar photon-dominated regions, PNe, etc).

We present 5-36~\mum\ spectra of 18 HAeBe stars, two intermediate-mass
T Tauri stars, HD 97300, and 51 Oph for a total of 22 sources observed with \textit{Spitzer}. 
This extends our original sample of HAeBe stars in Paper 1 by more than a factor of four, and we now
analyze HAeBe stars that do have the 10~\mum\ silicate emission feature. 
We present a detailed analysis of the PAH feature strengths and their ratios
in an attempt to infer their physical environments and
to test for trends in PAH emission when compared with other diagnostics of
proto-planetary disk evolution that have recently been helpful in understanding
the evolution of T Tauri disks \citep[e.g.][]{fur06}.

The major trends suggested in
Paper 1 persist in the present, larger sample. \cite{aa04}
noted in their study of 46 HAeBe stars that those with strong mid-IR 
excesses relative to near-IR excesses have stronger PAH emission
than those with weaker mid-IR excesses. This result supports a relation
between the detailed structure of PAH emission and the evolution of the disks. 
Recent studies by \cite{gra05} suggest that the strength of (UV-optical) photon-dominated
spectral features should correlate with overall disk geometry. Disks
that are flat and whose outer disks are therefore shadowed or illuminated at
grazing incidence should
have weaker H$_{2}$ line and PAH features relative to disks that are flared
and therefore well-illuminated by the photospheres of their host stars. These results
imply a strong correlation between PAH strength and disk structure,
which we can now test with our data.

\section{OBSERVATIONS AND ANALYSIS}  

\subsection{The Sample}

Table 1 summarizes the observations of our sample. SU Aur and HD 281789 
are intermediate-mass T Tauri
stars, which range in spectral class from K to late F (1.5-5.0 M$_\sun$). 
They are the evolutionary predecessors 
of HAeBe stars so their inclusion broadens our study to more primitive systems \citep{cal04}.
Both SU Aur and HD 281789 were included in the Taurus study of \cite{fur06},
in which the two stars appeared as clear outliers from the Class II \& III low-mass
YSOs that they focused on.
51 Oph is almost certainly not an HAeBe star \citep{thi05,ber07},
but we include it here because its mid-IR spectrum indicates an optically
thick, gas- {\it and} dust-rich disk with no measurable PAH emission and
because it appears in many previous HAeBe studies. We argue below that HD 97300 is probably
not a HAeBe star either. The remaining 18 targets are HAeBe stars.

\subsection{Observing Methods and Data Processing}

We observed all of these sources with the Infrared 
Spectrograph \citep[IRS,][]{hou04}\footnote{The IRS was a 
collaborative venture between Cornell University and
Ball Aerospace Corporation funded by NASA through the Jet Propulsion
Laboratory and the Ames Research Center.} on the {\it Spitzer Space 
Telescope}.  All but two of the sources were observed using 
the Short-Low (SL) module from 5 to 14~\mum, the Short-High 
(SH) module from 10 to 19~\mum, and the Long-High (LH) module 
from 19 to 36~\mum.  The remaining two were observed using SL 
and Long-Low (LL; 15-36~\mum).  SL and LL have resolving powers of 
R$\sim$90, while SH and LH have R$\sim$600.

Three of the spectra were obtained in simple staring mode,
while the rest were obtained as part of clustered 
observations, with three slit positions stepped
across the nominal source positions.  For these observations, 
we extracted a spectrum from 
the center position, unless the source was bright enough to 
saturate the detector array in SL, in which case we reconstructed 
the SL spectrum from the off-center positions.  
For sources saturated in SL, we used an average of
the two adjacent map positions and then scaled them with a multiplicative offset
until they matched-up with the short-high, long-high, or long-low spectra 
for the same source processed as
described below. Because the stars in question are point sources at the spatial resolution of the 
IRS, the line and continuum fluxes should scale together, but this method calls 
into question the absolute flux calibration of the 
short-wavelength portions of spectra that were saturated. 
However this scaling does not affect {\it ratios} of spectral
feature integrated fluxes that we derived from SL spectra
for use throughout this paper.
Readers using these data in other ways should be aware that
the photometry may be more uncertain than the signal-to-noise implies. Table 1 
summarizes how each source was observed and clearly identifies those
for which SL was saturated in the central map position. We present two of our
sources, HD 104237 and HD 100546, in SH and LH only since their
5--10~\mum\ continuum levels are above the saturation limits of the SL
module.

We started with the S14 pipeline output from the {\it Spitzer}
Science Center (SSC) and followed the standard calibration
method used at Cornell University, briefly summarized here. 
To remove the background emission 
from the SH and LH images, we subtracted images of the ``sky'' 
obtained by integrating on nearby offset fields.  For the SL
images, we subtracted images with the science target in the
other aperture, so that images with the target in SL order 2
served as sky images when the image was in SL order 1, and 
vice versa (i.e. aperture differences).  For LL, we 
differenced images with the source in the two nod positions in 
a given order (nod differences).  In addition to removing the 
background, this step also removes the majority of rogue 
pixels.  These pixels have temporarily high or low dark 
currents and are a significant source of non-gaussian noise. 
To correct for remaining rogue pixels and bad pixels 
identified in the SSC-provided mask images, we used 
imclean.pro\footnote{Available in IDL from the SSC as irsclean.pro.}

We extracted spectra from individual images using the profile,
ridge, and extract routines available as part of the {\it 
Spitzer} IRS Custom Extraction (SPICE) package.  We used 
standard stars to generate spectral corrections to calibrate 
the extracted spectra in flux-density units. For SL, HR 6348 
(K0 III) was the standard; for LL we used HR 6348, HD 166780 (K4 III), and 
HD 173511 (K5 III) as calibration standards.  For SH and LH, we 
used $\xi$ Dra (K2 III).  When combining the spectra from 
the two nod positions, we used the standard deviation at each 
wavelength to estimate the uncertainty.

The final ``stitch-and-trim'' step generates one continuous 
spectrum from the various segments.  We applied scalar 
multiplicative corrections to eliminate discontinuities, which 
result primarily from the fact that the target is better 
centered in the slit in some exposures than in others.  We scaled upwards to 
the best-centered aperture, which was usually LL order 2 or 
LH.  We combined bonus-order data in SL and LL with 
overlapping data in the other orders, then trimmed each spectral segment to
remove invalid data at the ends of the segments. Figure 1
presents the resulting spectra. 

Several sources have a choppy appearance in LH, which we
have called "corrugation". The effect is produced by a residual 
gradient signal on the IRS detector array (post flat field). The IRS team
is working to identify the physical source of this residual, which
generally appears at low light levels so usually only affects relatively faint spectra. 
We have removed 
the particularly bad corrugation in the spectrum of HD 97300 by fitting 
the residual signal between orders and subtracting it prior to extracting
the spectrum. The rest of the spectra 
are not as profoundly affected
and we have chosen to leave them alone since the corrugation effect does not
appear in the SL or SH modules where the PAH features that we are
studying appear.

\subsection{Overall Spectral Properties and Classification}

Figure 1 presents all of our spectra on the same axes with the flux scaled
arbitrarily to allow easy comparison. Figure 2 presents the data again with
each spectrum flux-calibrated and plotted on its own axes, $\lambda$ vs. F$_{\nu}$.
Blackbody emission from the central star and thermal emission from the circumstellar
disk dominate the broad-band shapes of the spectra while the detailed
structure comes from combinations of spectral features from PAHs and silicate dust grains.
In Figure 1 we have organized the spectra in groups according to the classification
scheme of \cite{kra02}: `SE' indicates the presence of silicate emission and `U' indicates
the presence of PAH emission (historically Unidentified InfraRed emission, 
hence the U). A lower-case
`u' indicates that the PAH emission feature contrast, relative to the continuum as measured
qualitatively by eye, is relatively weak.

The presence of high-contrast PAH emission in most of our sources
is striking particularly in the seven that have no silicate feature in the
10~\mum\ region (see Figure 1). The PAH feature centered at 7.7~\mum\ 
consists of two primary components, one
at 7.65~\mum\ and the other at 7.85~\mum\ \citep{coh89, 
bre89, pee02}. The 7.65~\mum\ feature tends to dominate the PAH spectra of reflection
nebulae and H II regions where the PAHs seem to have been heavily
processed. The 7.85~\mum\ feature dominates in many planetary nebulae
and objects evolving away from the asymptotic giant branch (AGB) where
the PAHs seem to be relatively fresh and unprocessed. We use
the PAH spectral classifications of \cite{pee02} who
studied 57 spectra of PAH sources obtained with the \iso\ SWS. 
Of those spectra, 42 belong to Class A,
which shows the PAH emission features at 6.2 and 7.7~\mum. 
They found that in spectra where the 7.85~\mum\ component dominated,
the 6.2~\mum\ PAH feature was shifted to 6.3~\mum. They used
these two characteristics as the basis for what they called Class
B PAH spectra, which are characteristic for 12 of their objects. Class C, 
where the peak emission in the
7--9~\mum\ range is shifted to the red of 8.0~\mum, accounts
for only 2 sources. Throughout the rest of this paper we refer to this feature complex as the 
7.7--8.2~\mum\ feature. One source in their sample had characteristics of both
Class A and B. We have applied this classification system to our larger HAeBe sample,
and we see a strong tendency for PAH emission from HAeBe to
to lie between Class B and C (Table 3). 

We have detected a strong 17~\mum\ feature in HD 97300. \cite{bei96} first 
identified the 17~\mum\ feature in \iso\ SWS 
spectra of Galactic PNe, attributing the feature to a PAH
in-plane deformation mode. Recently \cite{smi07} reported this feature
in \textit{Spitzer} IRS spectra of galaxies in the SINGS survey. They 
noted a strong correlation between the strengths of the 17~\mum\ and 11.3~\mum\ features,
concluding that the 17~\mum\ feature is associated with
PAHs in galaxies. We cannot verify such a correlation in our sample, since we have many
sources that excite the 11.3~\mum\ PAH features, but only one that excites the
17~\mum\ feature.  PAH emission observed in
other galaxies is dominated by interstellar photon-dominated regions, which
produce class A PAH spectra \citep{pee02}, while in our sample, HD 97300 is the only
Class A PAH spectrum.

In addition to the PAH spectral classification, it is useful to look for trends in the overall
shapes of the SEDs and the solid state features that are often prominent in HAeBe spectra.
\cite{mee01} and \cite{aa04} introduced a classification of the overall
shape of their \iso\ mid-IR SEDs for 46 HAeBe stars, including several
of those presented here. According to their classification, Group I SEDs
rise towards longer wavelengths when plotted $\lambda$ vs. $\lambda F_{\lambda}$. They suggested 
that Group I disks 
have a larger abundance of small dust grains heated in a radially flared disk. Group II
systems have bluer SEDs. Their conclusion is that Group II disks are probably more flattened due to 
settling of dust to the disk mid-plane. \cite{mee01} append the letter 'a' to indicate
the presence of solid state bands, most prominently the 10~\mum\ silicate
feature, or 'b' to indicate no solid state bands. We have included Meeus et al. group 
classifications with the individual spectra in Figure 2, and we have applied the Meeus et al. criterion
for stars in our sample that do not appear in their sample.

\subsection{Extraction of the PAH Emission Features}

We use a two-step algorithm to measure the integrated flux and
positions of the PAH features, following 
\cite{slo07}.  First, we fit a cubic spline to approximate and 
remove the underlying continuum from star, disk, and silicate 
dust grains.  This step removes the curvature under the broad 
PAH emission complex extending from 7.5 to 9.0~\mum.  We use two sets of 
spline anchor points, for spectra with and without silicate emission.  
Figure 3 illustrates this procedure
for spectra in our data set with different relative strengths of PAH and 
silicate emission.

The cubic spline fit leaves small residuals, which can result in slightly 
different continuum levels on the short- and long-wavelength sides of the
PAH features (Figure 4). When present these asymmetries
shift the apparent centers of the PAH features so we fit and subtract a line 
segment to each feature after fitting and removing the spline. 
Table 2 gives the 
wavelengths used to fit these segments.  We fit the continuum 
between the blue and red pairs, and integrate the feature 
between them to obtain our integrated fluxes.  Figure 4 illustrates this process of measuring
the PAH features in the spectrum of HD 135344.\footnote{SIMBAD now
refers to this source as HD 135344B. In any case it is still SAO
206462.} For each PAH feature, we measure the integrated flux above the fitted line segment
and we measure the central wavelength.  

To determine the central wavelengths of the PAH features from the continuum-subtratcted spectra
we first measure the feature integrated flux, F. We then integrate the flux in each feature starting from the
continuum on the short-wavelength side and stopping at the wavelength corresponding to 0.5F. We
integrate the flux in the same feature from the continuum on the long-wavelength side of the feature and
stopping at the wavelength corresponding to 0.5F. We define the central wavelength as the average of 
these two wavelengths. The wavelength corresponding to the integrated flux median 
of a feature is not necessarily the same as its center wavelength 
because there is some uncertainty in locating the wavelengths at
which the flux begins to rise above the noise on either side of the feature. 
We estimate the uncertainties in the central 
wavelengths by taking the wavelengths corresponding to 
$F_{\nu}+\sigma_{F_{\nu}}$ and $F_{\nu}-\sigma_{F_{\nu}}$, subtracting them,
and dividing by 2. Here $F_{\nu}$ is the flux density at the central wavelength and
$\sigma_{F_{\nu}}$ is the uncertainty in $F_{\nu}$. Figure 5 presents the fully continuum-subtracted
spectra for all sources in our sample organized roughly in order of
decreasing total PAH integrated flux.

As \cite{slo07} demonstrated,
the wavelengths of PAH features shift as the temperature
of the radiation field varies.  The broad PAH features are 
generally thought to be combinations of narrow features
\citep[e.g.][]{bt05}, and the central wavelength is more
sensitive to variations between the strengths of various
components than the peak wavelength, which will remain at
the peak of the strongest feature despite variations in the
strengths of any secondary components. We present central
wavelengths in Table 3 and feature integrated fluxes in Table
4. Note that we list upper limits equal to $\sigma_{F_{\nu}}$ 
in cases where the signal-to-noise
ratio is S/N$<$1.

For spectra with a PAH emission feature at 6.0~\mum\ measured
to have a S/N of two or more, we remove this feature before 
determining the strength and position of the 6.2~\mum\ 
feature.  Similarly, we remove the 8.6~\mum\ feature before 
measuring the 7.7--8.2~\mum\ feature.

The weakest continuum-subtracted PAH spectra 
(HD 32509, 51 Oph, HD 152404, HD 144432, and HD 31648)  are for stars with
strong silicate dust features at 10~\mum. Because the PAH features for
these sources are both intrinsically weak and superposed on a continuum
that is steep and varying in slope (e.g. the 10~\mum\ feature itself), we have low
confidence in the accuracy of our extracted PAH center wavelengths.  
We have therefore omitted these five sources from the feature wavelength 
analysis below (Figures 6-9). We use their integrated PAH flux in our comparison of PAH
emission to SED shape as inferred from continuum color 
indices (see further discussion in Section 3.2.1), 
since that analysis relies on a more qualitative treatment of overall PAH emission
and does not include PAH feature wavelengths. The five sources 
in question are identified with asterisks
in Figure 5. For HD 104237 and HD 100546, 
we cannot extract PAH features because we do not
have usable SL spectra for those two stars; both are too bright for the 
IRS SL module. We present their
SH and LH spectra in Figure 2 and their continuum-subtracted SH spectra
in Figure 5, but do not include them in the analysis of PAH feature positions
and integrated fluxes below.

\subsection{PAH Flux Ratios}

The 6.2 and 7.7--8.2~\mum\ features both arise from C$-$C 
bonds \citep{atb85}, which are enhanced in ionized PAHs, while the 
11.3~\mum\ feature, which arises from a C$-$H bending mode, is stronger 
in neutral PAHs  \citep{ahs99,dra03,dra07}.  Plotting the ratio of either of the C$-$C 
modes over the 11.3~\mum\ integrated flux therefore indicates the degree of 
ionization in the PAHs.  In Paper 1 we examined 
these flux ratios in a sample of only four Herbig AeBe stars 
and found that while both ratios are supposed to measure 
ionization, we could not identify an obvious trend with a sample
of only four stars.

The 11.3 and 12.7~\mum\ features are both C-H out-of-plane
bending modes, but the 11.3~\mum\ feature arises from carbon 
rings with only one adjacent hydrogen (the solo mode), while
the 12.7~\mum\ feature arises from rings with three adjacent
hydrogens (the trio mode).  \cite{hon01} investigated the
ratio of these two features and found that it correlated with
the PAH ionization ratio (as measured from the ratio of the 
7.7--8.2 and 11.3~\mum\ features).  They noted that PAHs with
long, straight edges would have more emission at 11.3~\mum\
than at 12.7~\mum, while PAHs whose edges had been broken up
by radiative processing would have higher 12.7/11.3~\mum\
ratios.  We found a tentative 
correlation in Paper 1 for our small HAeBe sample that supported the
conclusions  of \cite{hon01}.

We have plotted these integrated flux ratios for our larger sample of
Herbig AeBe stars in Figure 6, along with the class C PAH sources and 
the class A and B prototypes discussed by \cite{slo07}. The Class A prototype is
NGC 1333 SVS 3 (a reflection nebula), the Class B prototype is HD 44179 
(a post-AGB source). The Class C sources are: SU Aur (an intermediate-mass T Tau
star in our sample), IRAS 13416-6243 and AFGL 2688 (both post-AGB), and two red
giant stars HD 233517 and HD 100764.
The top panel plots the 6.2/11.3~\mum\ ratio vs. the 
7.7--8.2/11.3~\mum\ ratio. The data show a 
rough increase in one flux ratio as the other increases,
but with substantial scatter. The variation in the relative
strengths of the 6.2 and 7.7--8.2~\mum\ features is enough to
affect estimates of the ionization ratio using only one of
these two bands.  Including the point for HD 141569 makes it
tempting to report a correlation between the two, but we
hesitate to draw any conclusions without data to fill the
gap between it and the rest of the sample. The bottom panel of 
Figure 6 plots the relative strength of
the trio and solo C$-$H modes vs. ionization ratio, F$_{12.7}$/F$_{11.3}$.  
A trend of increasing F$_{12.7}$/F$_{11.3}$ with increasing ionization is 
evident.

Figure 7 illustrates that the 12.7/11.3~\mum\ integrated flux ratio 
varies relatively smoothly with the temperature of the 
central source.  This figure (and the bottom panel of 
Figure 6) exclude a point at F$_{12.7}$/F$_{11.3}$ = 1.4 from
AFGL 2688, one of the class C PAH sources.  This spectrum and
the spectrum of IRAS 13416$-$6243 have PAH features with very
low contrast with respect to the strong red continuum, making
our measured feature sensitive to any artifacts introduced 
during the spline-fitting and extraction \citep[see][for more details]{slo07}.  
IRAS 13416 is one of the two remaining
outliers (at T=5440 K), for the same reason.  The remaining
outlier is HD 141569, which is highly ionized for its 
temperature (Figure 6) and appears to be more heavily processed
than otherwise similar objects.

\section{DISCUSSION} 

Since PAH emission is present in so many HAeBe stars and, when present,
may play a significant role in the energy balance of the circumstellar
material \citep{kam04}, we should ask both what the PAH emission tells us about
circumstellar disks orbiting intermediate-mass stars and what the
environment around these stars can help us learn about PAHs as
diagnostics in other astrophysical environments. In Paper
1 we concentrated primarily on the latter analysis because of the
small size of our HAeBe sample at the time. In the present study we expand
the detailed analysis of mid-IR PAH emission from HAeBe stars that
we started in Paper 1, including comparisons of the spectra to preliminary
models of Ae star-disk systems. We find that the larger sample reveals some
trends that will help with both questions. A detailed discussion of the 
physical nature of circumstellar disks necessarily 
depends on accurate measurements or derivations of the physical characteristics
of the host stars, which we summarize in Table 5.

\subsection{Physical Characteristics of the PAH Molecules and Their Environments}

In Paper 1 we noted that the four HAeBe stars we studied had PAH spectra
with their 6.2~\mum\ and 7.7--8.2~\mum\ features shifted to the red \citep[e.g.][Class
B or Class B/C]{pee02}
relative to PAH spectra from HII regions. Adding more stars to the sample increases our
confidence in this conclusion. The Peeters et al. sample
included only two isolated HAeBe stars (HD 100546 and HD 179218),
but their classifications indicated that the degree of processing
of the PAH molecules is less in HAeBe disks than in
interstellar PDRs. That trend is much more obvious in the present HAeBe
sample.

Photo-processing of PAHs by the stellar radiation field is evident in our analysis
of the spectra of HAeBe stars. Figures 8 and 9 show that all of the PAH 
features in the 6--13~\mum\ range have their centers
shifted to redder wavelengths as the stellar effective temperature decreases.
Note that the vertical scatter in the data is real, not statistical, in the sense that earlier episodes
of PAH processing (e.g. prior to the current stellar radiation field) have already altered the
emission feature centers to varying degrees \citep{slo07}. Nevertheless a clear trend is evident in the data.
This trend strengthens the tentative conclusion in Paper 1 that the shifts in PAH 
feature wavelengths indicate the degree to which the molecules have been exposed to the
stellar radiation fields and therefore how processed they are (i.e. ionized, fragmented, and/or
dissociated) for a given star-disk system. The Peeters et al. classifications 
appear to be tracing the extent to which
PAH molecules have been processed by their environments, particularly
the stellar radiation field, with Class A representing the most processed and
Class C the least processed PAHs. Isolated HAeBe stars occupy a narrow range
spanning Classes B and C.

The HAeBe PAH spectra resemble those of post-AGB objects
in which the PAH molecules are relatively \textit{un}processed (i.e.
newly formed). In the case of HAeBe stars, since there is no clear
formation process for PAHs in the stars or the disks, it is more likely that the
PAHs appear less processed because they are newly released into the
stellar radiation field from some primordial reservoir. The source of 
PAHs may be evaporation of the icy
mantles on dust grains formed earlier in the proto-{\it stellar} material and
containing pristine PAH molecules (e.g. preserved from the pre-stellar ISM).
This implies that PAHs in HAeBe disks are constantly
being destroyed \textit{and} replenished from the reservoir by the stellar UV-optical radiation
field. It would also explain the strong PAH sources
that completely lack 10~\mum\ features commonly associated with a
large component of warm, relatively small silicate dust grains in disks.
Icy mantles on larger dust grains can evaporate, releasing PAHs 
in the absence of small silicate grains. We suggest that disks with
this "PAH no-silicate" spectrum (in our sample, HD 34282, HD 97048,
HD 97300, HD 100453, HD 135344, HD 141569, and HD 169142) have
a large inner cavities that are cleared of dust out to radii of at least a few AU, i.e. disks with inner
holes. All of the stars in our sample that have this distinctive spectrum, except
HD 97300, share an overall SED structure resembling the transitional disks recently
identified and studied in T Tauri systems \citep[e.g.][]{uch04,dal05,cal05,esp07,bro07}.
Since PAH molecules are stochastically heated by individual optical-UV photons
they can emit despite being located far from the star while the dust in the transitional disks is too far, therefore too cold, to emit the 10~\mum\ feature. With the exception of HD 141569, the "PAH no-silicate" systems have strong near-infrared excesses. The NIR excess in these systems may be a combination of
strong NIR line emission (e.g. warm molecular gas emission from the inner cavity and 
recombination emission from gas accretion flows) and black-body emission from small
bodies (e.g. very large dust particles, planetesimals, asteroids) orbiting in the inner disk region.



\subsection{Disk Geometry and the Locations of the Emitting PAH Molecules}

\cite{mee01} suggested that HAeBe stars with redder (Group I) SEDs
have a long-wavelength component indicating smaller dust grains
that are located in a relatively warm, radially flared region of the disks. Their
Group II contains SEDs that lack the warm, small-grain component. They interpreted 
Group II SEDs as indicating a more flattened disk containing mostly larger dust grains
that have settled to the disk mid-plane, whereas Group I SEDs represent
disks with a substantial radial flare. Subsequent modeling by \cite{dul03} 
have supported these conclusions. \cite{mee01} and \cite{aa04} also noticed a tendency
for the Group I sources to be among the $~50\%$ in which they detected PAH features
in \iso\ SWS spectra. For the sources with stronger PAH emission, our results are
consistent with \cite{aa04}; the strongest PAH sources are in fact all Group I. However the higher 
sensitivity of the {\it Spitzer} IRS has revealed that some Group I sources have weak
PAH emission and that many Group II sources that were tentative ISO SWS detections in only the
6.2~\mum\ feature are PAH emitters in all of the mid-IR PAH features (e.g. HD 141569, HD 142666,
HD 35187, HD 31648, HD 144432). Thus, although it does seem that the PAH emission comes 
from the illuminated surfaces of flared disks, we suggest that such emission is {\it not} an 
unambiguous tracer of flared geometry in HAeBe disks.

\cite{dou07} recently detected the flared disk around HD 97048 in a mid-IR direct imaging study.
The flaring disk is evident in the image, and the source is more extended in the
11.3~\mum\ PAH feature than in the adjacent continuum. \cite{gra05} reported that mid-IR PAH
emission, in particular the 6.2~\mum\ feature, correlates with optical scattered light
visibility of the inner disks (r$\sim$50-70 AU) of some HAeBe stars. Since
the PAH features and H$_{2}$ line emission are often strong where the optical-UV radiation
field is strong (e.g. interstellar photon-dominated regions), they should trace material 
that is far enough above the disk mid-plane to be directly illuminated by the star. \cite{gra05}
therefore suggest that both the optical visibility of the disk and the
mid-IR PAH emission should correlate with flaring and anti-correlate with
dust settling to the disk mid-planes. However, four of the STIS scattered light {\it non-detections} 
from \cite{gra05} are in our sample (HD 31648, HD 104237, HD 135344, HD 142666). 
All but HD 104237 clearly have PAH emission in our spectra. These apparently optically dark 
disks have optically visible stars so their
disk surfaces cannot appear dim to us simply because we are seeing them
nearly edge-on. In fact their inclination angles range from 38\arcdeg\ to 
55\arcdeg\ (Table 5). Apparently disks with relatively little scattered optical light are still illuminated
enough to excite PAH emission, or the PAH emission comes from the inner disk and {\it not} from
the atmospheres of the outer disks.

Although it is tempting to use the PAH emission strength as a 
preliminary diagnostic of overall disk geometry, especially for those sources whose disks
are unresolved or marginally resolved in visible, near-IR, and mid-IR images, we 
suggest that some caution
is indicated. The question remains as to whether large-scale radial flaring dominates
or whether shadowing of the outer disk by 
structures in the inner disk may play a role. The former possibility implies
that flared disks should be strong PAH emitters unless we happen to view 
them edge-on. The latter possibility implies that illumination of even highly flared disks can
nevertheless be partially and/or temporarily attenuated by structures in the inner disk--within
a few AU of the star--and therefore produce weak or non-PAH emitters as well as 
weak or non-detections in optical/NIR scattered light observations.

\subsubsection{SED Shape and PAH Emission}

The geometry of the disks alters their SED significantly and there have been
many efforts to quantify this relationship for circumstellar accretion disks,
including those around HAeBe stars, both empirically 
(e.g. \cite{mee01}) and with detailed models \citep[e.g.][]{dul03,dul04,dal01,dal05,dal06}.
Recently \cite{fur06} and \cite{wat07} have combined these approaches
to study disks around low-mass young stellar objects in Taurus. They used continuum color indices 
derived from {\it Spitzer} IRS spectra to characterize the SED shape for comparison to 
grids of accretion disk models by \cite{dal06}. They defined the color 
indices by differencing fluxes in two narrow wavelength bins located between the prominent solid state features,
essentially providing a measure of the slope of the SED between the two wavelength
bins. The differences in flux ($\lambda F_{\lambda}$) from one bin to the next (e.g. from a to b), 
expressed as ratios of their base-ten logarithms, form the color index, $n_{a-b}$:


$n_{a--b}=\log(\lambda_{a} F_{\lambda a}/\lambda_{b} F_{\lambda b})/ \log(\lambda_{a}/\lambda_{b})$.


We have chosen spectral bins for our indices that avoid both the solid state features
and the PAH features that are common in HAeBe spectra: 
5.4--6.0~\mum, 13.0--14.0~\mum, and 29.75--31.75~\mum,
labeled 6, 13, and 30~\mum\ respectively. Following \cite{fur06}, we have analyzed
our sample stars using $n_{6--13}$ vs. $n_{13--30}$ color-color diagrams.

\cite{dal06} found that model star-disk systems changed location on such color-color 
diagrams extracted from grids of model SEDs in which degrees of dust settling 
and disk inclination were allowed to vary. 
\cite{fur06} and \cite{wat07} found that plotting their samples of low-mass YSOs 
on the same color-color diagrams indicated a wide range of dust grain sizes, which
they interpreted as degrees of dust settling to the disk mid-plane. D'Alessio and
collaborators have begun generating similar models for Herbig Ae stars and we can
now begin a similar analysis for intermediate-mass star-disk systems. Since these models
are still in a preliminary stage of development, in this paper we will use them only to infer schematic trends
in disk physical characteristics indicated in continuum color-color diagrams. More detailed 
predictions and estimates of disk properties for individual HAeBe 
systems will be the focus of a forthcoming study.

\subsubsection{Model Physics}

The fundamental physics of the disk model we use is described in \cite{dal98,dal99}. 
We assume a steady
state, axisymmetric and geometrically thin disk. Its mass accretion rate
($\dot{M}$) is taken to be uniform (spatially constant). We decouple
the equations for the disk vertical and radial structure since
the disk is geometrically thin. In the vertical direction the disk is in
hydrostatic equilibrium with the energy being transported by
radiation, turbulent conduction, and convection; the latter mechanism is
active only where the convective instability criterion is
satisfied. In the radial direction, the disk is assumed to be in
centrifugal balance in the potential well of the star, i.e., a
Keplerian disk. The primary disk heating sources are viscous
dissipation, with the viscosity coefficient given by the $\alpha$
prescription \citep{sha73}, and stellar irradiation. In
order to calculate the transfer of disk and stellar radiation, we use
mean opacities in which the average of the monochromatic
opacity is calculated using, as weighting
functions, either the Planck function (or its derivative with respect to
temperature) evaluated at the local kinetic temperature or the star
radiation temperature, respectively.  For simulating the effect of
dust settling, we assume that the disk has two populations of dust grains,
with different grain-size and spatial distributions. In the disk atmosphere
there are small grains, depleted with respect to the standard dust to
gas mass ratio \citep{dal06}. Closer to the mid-plane,
there are larger dust grains, with an enhanced dust to gas mass ratio,
which accounts for the mass in grains that have disappeared from the
upper layers.  For simplicity, we assume that the settling is only in the
vertical direction.  The boundary conditions for the transfer
equations are given in the "irradiation" surface, defined as the
surface where the mean radial optical depth to the stellar radiation
is unity. The larger the depletion in the upper layers, the smaller
the height of the irradiation surface, i.e., the smaller the degree of
flaring of the disk. The flatter the disk, the smaller the fraction of stellar
radiative flux intercepted by its surface.

The disk inner boundary is assumed to be the radius where optically
thin dust, heated by the star, reaches its sublimation
temperature. Assuming that silicates comprise the dust component with
largest abundance, sublimated at the highest
temperature, we adopt a dust sublimation temperature of 1400 K.  The disk inner wall
model is described by \cite{muz03} and \cite{dal05}. It is assumed to 
be a plane-parallel atmosphere, with
constant temperature in the vertical direction, heated by the
star in the radial direction. The  temperature of this atmosphere
is calculated as a function of the  mean optical depth in a direction parallel
to the disk mid-plane (radial, in cylindrical coordinates).
The vertical height of the wall is roughly estimated as the height
where the mean radial optical depth to the stellar radiation is unity,
which depends on the opacity and density at the wall. We assume the
wall is in vertical hydrostatic equilibrium with gas  in LTE and
the same dust properties as in
the disk atmosphere. The SED of the wall and the SED of the disk are calculated 
separately.

All of our disk models are flared to varying degrees. 
Being 1+1D (i.e., with the vertical and radial structure
calculated separately), these models do not account for the possibility
of shadowing. However, the scale height of the wall and the scale
height of the flared disk at the same radius are very similar, mostly
because the mid-plane of the disk is heated by viscous dissipation,
which is not included in our wall model. This heating mechanism seems
to compensate for the difference in the incident stellar flux between
the wall and the disk. Thus, our disk models are not puffed-up in the
way described by \cite{dul01}.  When the wall has dust
settled towards its mid-plane, its effective height (i.e.,
the size of its emitting area) is smaller than the the wall with no
settling. Thus, the contribution to the SED of the former is
smaller than the contribution of the later. The SEDs of both kinds of
models are added to the same SED of the rest of the disk, from which
the contribution to the emission of its inner boundary has been
removed.

\subsubsection{The distribution of PAH emission with color index}

In Figure 10 we present n$_{6-13}$ vs. n$_{13-30}$
color-color diagrams derived from a grid of model disk systems for stellar spectral types
A2 and A6, as well as varying inclination angle (40\arcdeg - 70\arcdeg), mass accretion rates 
($10^{-8}$ and $10^{-9} M_{\sun}/yr$), age (1 Myr - 10 Myr), and 
degree of atmospheric dust grain depletion ($\epsilon$=0.001 - 1.0). The depletion factor, $\epsilon$, 
varies from 1.0 to 0.001 and represents the dust-to-gas mass ratio for disk atmospheric silicate
dust grains relative to the solar dust-to-gas mass ratio \citep{dal06}. Thus $\epsilon$ 
is a measure of the depletion of atmospheric grains as they settle to the disk midplane. 
Decreasing $\epsilon$ is an indicator
of overall disk settling from flared to more flattened geometry. The more flattened 
disks will tend to occupy
the bluer (lower left) region of the color-color diagram in Figure 10, while more flared
disks will be redder and located towards the upper right. Disk inclination increases right to
left while stellar age increases from lower right to upper left. 

Figure 10a contains models for mass accretion rates of $10^{-8} M_{\sun}/yr$ (top panel)
and $10^{-9} M_{\sun}/yr$ (bottom panel)
in which the dust in the inner wall of the disk has settled. Figure 10b, with the same 
accretion rates in the top and bottom panels,
contains models for which dust in the inner wall has not settled. The inner wall is located at the inner
radius of the disk.
We have over-plotted our observed color indices for comparison with the models.
The models and observed colors overlap, but the observed HAeBe stars 
cover a larger range of parameter space than the current grid of models.

We plot our star sample on an identical color-color diagram in Figure 11 with the
PAH flux (Figures 11a and 11b) indicated in the figure by the size of the hexagonal symbols that are
centered on the star colors. We calculated the total PAH flux for each star by summing the integrated
fluxes of the 6.2, 7.7--8.2~\mum\, 11.3, and 12.7~\mum\ features and dividing by the square of the
stellar distance normalized to 100 pc so that the symbol sizes in Figure 11a represent the
intrinsic total PAH emission for each star. Because the 11.3~\mum\ PAH feature coincides with the
11.1~\mum\ feature of crystalline silicates (olivine) and since both can be strong in 
HAeBe stars (e.g HD 144432, Figure 2), we have also computed luminosities for the 6.2~\mum\
feature alone (Figure 11b). The 6.2~\mum\ feature is not near other
PAH or silicate features so it serves as an independent indicator of PAH emission strength.
A comparison of Figures 11a and 11b shows that they are not identical, but the 
feature that we extracted at 11.3~\mum\ seems to be dominated by the PAH emission. Note that our continuous color indices are consistent with the
discrete Group designations of \cite{mee01}. Group I disks are flared and Group II disks are more flattened.
Nevertheless, Group I  disks have a wide range of PAH integrated fluxes.

\subsubsection{Notes on HAeBe stars that do not overlap the models}

In our analysis of color indices in the previous section we noted that our data have a larger
range of colors than the present grid of models. Twelve stars have colors outside the range
of the model grid: HD 32509, HD 34282, HD 281789, HD 152404, HD 169142,
HD 97300, HD 135344, HD 139619, HD 141569, HD 169142, 51 Oph, and Elias 3-1. Two of these, 51 Oph and 
HD 97300 are probably not HAeBe stars. The remaining
ten stars include HD 32509, which has NO measurable PAH emission, and five Group I stars with relatively weak PAH emission. Furthermore, {\it all} of the seven sources in our sample that have 
PAH emission, but no 10~\mum\ silicate emission feature, are in this "outlier" sub-sample. 

HD 141569 is a more evolved system with substantial clearing of material in its disk \citep{mer04,
aug99, wei99} . If the inner disk structure allows stellar radiation farther into the disk where PAHs
may be released from icy mantles on large grains, this could explain why HD 141569 is by far the most ionized PAH 
spectrum in our sample. HD 169142 appears to be in an epoch of central clearing of the disk and may resemble 
Vega in an earlier epoch of its disk evolution \citep{gra07}. The disk orbiting HD 100453  seems to be dominated 
by large dust grains \citep{vdb04}.

HD 97300 is a B9 star located in the Chamelion I cloud that is close to the zero-age main sequence and has no near-IR excess emission or H$\alpha$ emission \citep{sie00}. \cite{sie98} report that the PAH 
spectrum is a mixture of a small amount of emission from a circumstellar disk and a large amount from the very 
bright, ring-shaped reflection nebula that HD 97300 illuminates. Although HD 97300 has been included in past 
studies of Herbig AeBe stars, we suggest that its Herbig Be classification may be in error since it is the only Class A PAH spectrum in our sample and the only source in our sample with emission in the 17~\mum\ feature. Certainly the fact that it lacks both H$\alpha$ emission and any near-IR excess suggests  a more evolved system that has ceased active accretion. The PAH spectrum of HD 97300 is probably dominated by 
interstellar PAHs rather than PAHs in a circumstellar disk.

\subsection{What does PAH emission tell us about HAeBe disks?}

At first glance a weak correlation of PAH luminosity with color index is evident in Figure 11a.
Although there are no strong HAeBe PAH emitters that have colors indicating highly settled disks, 
there are six relatively weak PAH sources that have colors indicating that they should have flared 
disks: HD 34282, HD 100453, HD 35344, HD 139614, HD 141569, and SU Aur. The Meeus Group classifications and the spectral indices are consistent in their
segregation of the stars in Figure 11, which we illustrate schematically with a dashed line
that roughly separates the Group I and Group II sources in our sample. We use Figure 12 to illustrate the
frequency of Groups I and II as a function of the n$_{13-30}$ spectral index (the vertical axis
in Figure 11). The Meeus Groups form a bimodal distribution in our sample that is consistent
with the interpretation that those Groups distinguish between flared and flattened (or non-settled and
settled) disk geometries.
Nevertheless, we see overlap in Figures 11 and 12; there are relatively weak PAH emitters that should
have radially flared disk geometries according to both discrete (Meeus Group) and continuous
(spectral index) SED classifications. Why do some apparently flared disks have weak PAH
emission?

We suggest several possible explanations for the absence of strong PAH emission in 
radially flared disks, which should have a large surface area directly illuminated by the star: (1) the 
abundance of PAHs in the disk atmospheres of those systems (either free-floating or locked-up in icy dust grain mantles) is intrinsically low; (2) PAHs are abundant in the disk atmospheres, but a strong stellar UV flux has destroyed most of the PAHs in the emitting region of the disk; (3) we are viewing those 
systems edge-on or nearly so; or (4) the inner walls of the disks are attenuating or softening the stellar illumination of a flared outer disk as suggested by \cite{aa04} and explored in detail more recently by \cite{ise06}. We now explore these hypotheses in turn.

{\it PAH Abundance.} If, as we suggested earlier, most of the emitting PAHs in HAeBe disks are released from icy mantles and if the PAHs are then destroyed by
the photo-processing that makes them emit, it is possible that the icy reservoir might eventually
be exhausted. In this scenario the radially flared disks with relatively weak PAH emission may simply have
evaporated their PAH reservoirs more thoroughly than the stronger PAH emitters. With the exception of
SU Aur, an intermediate-mass T Tauri star, the flared disks with weak PAH emission all have SEDs that
resemble transitional disks. They may be the most evolved disk systems in our sample.

{\it Destruction of PAHs.} We cannot blame a strong UV field destroying PAHs since some of the hottest stars in our sample excite very strong PAH emission (e.g. Elias 3-1, AB Aur, \& HD 97300).
We illustrate this in Figure 11c, where  the symbol sizes are proportional to the total PAH flux 
density ($\Sigma F_{PAH}$) divided by the stellar bolometric luminosity ($L_*$) for each source. In Figure 
11d, symbol sizes are proportional only to the stellar bolometric luminosity. We calculated 
bolometric luminosities using published values of the apparent visual magnitudes, distances, and 
extinctions for our sample stars (Table 5). Bolometric corrections are for luminosity class
V \citep{kh95}. There is no indication in Figures 11c-d that stellar luminosity correlates inversely
with PAH intensity. There is some indication that the opposite is true, but not uniformly in our 
sample (Figure 11d). Finally, as a sub-sample the "weak" PAH Group I sources are not uniformly the most 
processed nor are they all the most ionized sources in our sample. We therefore conclude that 
differences in the stellar radiation field from system to system alone cannot account for the variations in PAH emission.

{\it Disk Inclination.} For sources with measured inclinations we see no indication that
more inclined disks (large $i$, see Table 5) have weaker PAH emission. None of the stars have silicate 
absorption features at 10~\mum\ and those with
the weakest silicate emission have very strong contrast in the PAH features.
Thus inclination of the disks along our line of sight cannot  explain the variation in PAH luminosity among 
the redder disks. Table 5 also lists measured extinctions towards most of our sample stars, which
show no trend towards decreased PAH emission for more highly extinguished sources. In fact, none of
our sources, not even Elias 3-1 with A$_v$=4.05, has enough extinction to significantly alter its mid-IR SED. 
The bottom line is that there are highly flared disks nearly facing us that are nevertheless weak PAH 
emitters. The models summarized in Figure 10 indicate that inclination should affect
the mid-IR colors of our sources and though we see a range of colors, we see no trend for PAH
emission to weaken as colors change due to disk inclination (e.g. from lower right to upper left in Figures
10a and 10b).

{\it Attenuated Radiation or Shadowing by the Inner Disk.} The spatial distribution of PAH molecules in proto-planetary disk systems is a subject of current discussion. Since there is no clear formation path for PAHs in proto-planetary disks, the PAHs are likely primordial in the sense that they were present in the ISM 
before the star and disk formed \citep{ll03}.  \cite{aa04} concluded that the PAH emission must originate 
from the denser environment of a disk rather than from a large diffuse halo or envelope around the 
star. More recently \cite{vis07} have modeled the chemistry and infrared emission from
PAHs around HAeBe stars. After comparing their models to observational studies of
extended PAH emission around HAeBe stars \citep[e.g.][]{gee05,gee06,hab05,vdb04} they conclude
that PAH molecules with sizes of $\sim$100 carbon atoms are responsible for most of the 
observed emission. PAH molecules that big can survive the stellar radiation fields
of intermediate-mass stars down to a minimum radius of $\sim$5 AU. Thus, \cite{vis07}
predict that about 60\% of the PAH emission in the 6--13~\mum\ range originates within
100 AU of the star.  

For disks that are no longer accreting material onto their stars, a puffed-up inner wall in the flared Group I disks 
may attenuate or soften the stellar radiation field illuminating some or all of the outer disk.
This may be a relatively short-lived stage in what appear to be 
disks in transition from flared to flattened. If planet formation clears an inner hole, then the gas in the inner wall could be vertically inflated relatively quickly (and to varying degrees depending
on the system) by photoelectrons ejected by UV photons from PAHs and small dust grains. Thus for a short time, prior to the cooling and settling of the outer disk, there can be both a high inner wall and a
radially flared outer disk. As the grain growth and disk settling proceed, the flare of the disk may decrease 
until the PAH-rich parts of the outer disk settle into the shadow of the inner disk wall. Thus redder disks that 
are weak PAH emitters may be in an intermediate or transitional stage of disk evolution. We note that this suggested
scenario is beyond the scope of our current grid of disk models to predict since those models contain no treatment of shadowing in the disks.

\section{SUMMARY AND CONCLUSIONS}  

We have presented \textit{Spitzer} IRS spectroscopy of a sample
of 22 targets including 18 Herbig Ae/Be stars and two intermediate-mass T Tauri stars
whose spectra show strong PAH emission. The
absence of silicate emission in seven of the sources has enabled a
detailed analysis---independent of mineralogy estimates or specific
disk models--of the relative strengths and positions of PAH emission
features in the 5--14~\mum\ region. For the remaining stars
that do have solid state emission features, we have run the same PAH
spectral analysis by fitting and subtracting dust continuum and dust
emission features with a model-independent method, isolating the 
PAH spectrum. This relatively
large sample and the detailed analysis afforded by high signal-to-noise
IRS spectra led us to the following conclusions:

\begin{itemize}

\item PAHs exposed to stronger
stellar radiation fields may become shattered and develop more jagged
edges so that the 11.3~\mum\ mono C-H bending mode is attenuated with respect
to the 12.7~\mum\ trio C--H bending mode. PAH ionization fraction, inferred
from the PAH F$_{7.7-8.2}$/F$_{11.3}$ ratio, is related to the structure (and
possibly the size) of the PAH molecules. 

\item The center wavelengths of the 6.2, 7.7--8.2, 8.6, and 11.3~\mum\ emission features 
change as the stellar effective temperature changes. We confirm 
our previous finding that the center wavelengths of the
PAH emission features increase with decreasing effective
temperature of the host star, now with a much larger sample of stars. We conclude
that hotter stars photo-process the PAHs more than cooler stars do and that
the degree of processing alters the PAH spectrum measurably.

\item Our sample shows spectral characteristics generally consistent with
the boundary between Class B or C PAH emission sources defined by \cite{pee02}, viz.
a 6.2~\mum\ feature shifted to 6.3~\mum\ and 7.7--8.2~\mum\ emission
dominated by the component at 7.9~\mum. However, in some of our
sources, the 7.7--8.2~\mum\ feature center is shifted closer to 8.0~\mum, as
in the spectrum of HD 100546, a HAeBe star observed by the SWS on
\iso\ and included in our sample. It is remarkable that HAeBe stars
as a class have such a variety of PAH luminosities and ionization
fractions, but such a narrow range of PAH spectral classifications. This may
indicate that the PAH molecules are altered by the same physical processes
in the disks, perhaps unique to HAeBe stars.
In any case we propose that the HAeBe stars comprise a distinct PAH spectral
group of their own. HAeBe PAH spectra indicate very little processing of the
PAH molecules by the stellar radiation field, relative to other types of
PAH source (e.g. HII regions, PDRs, and PNe).

\item The other Peeters Class B objects are post-AGB and
that implies that the PAH are less processed and therefore relatively new to the scene in 
Herbig Ae/Be disks. This is consistent with the suggestion that the PAHs are introduced
at a stage of disk evolution when icy mantles evaporate from large dust grain
surfaces. If this is the case, then the PAHs near HAeBe stars are being
constantly destroyed and replenished by interactions with the stellar radiation
field since we do not observe HAeBe systems with the (apparently more processed) 
Class A spectra.

\item We have detected the 17~\mum\ PAH complex in one of our sample
HAeBe stars, HD 97300, which stands apart form the rest of our sample as the
only class A (e.g. highly photo-processed) PAH spectrum and is likely not a Herbig
Ae/Be star. The 17~\mum\ feature 
is commonly observed in the Galactic ISM and in other galaxies where strong UV fields excite extensive 
PDRs. HD 97300 is also a Meeus Group IIb source having no warm small grain component in its 
SED and no silicate features at 10~\mum\ or longward of 20~\mum. We suggest that, whatever
the classification of HD 97300, it's spectrum is dominated by PAHs in the surrounding interstellar
gas and not in a circumstellar disk.

\item Our sample includes both flared (Meeus et al. 2001, Group I) and flattened/settled (Group
II) systems, but the overall PAH emission does not decrease 
monotonically with indicators of depletion of small dust grains and disk settling, as
one would expect if the PAH emission only originates in an optically
and geometrically thin surface layer of a radially flared disk.

\item Taking into account estimates of
disk inclination for the sources in our sample does not
uncover a direct correlation of weak PAH emission with indicators of nearly edge-on disk
geometry. Nor can we explain the anomaly by invoking strong stellar radiation fields
to destroy or process the PAHs so that they emit less light in some disks.

\item Differences in PAH abundance in the illuminated parts of HAeBe disks and/or evolution of the inner
disk scale height appear to be possibilities for explaining highly flared disks that have relatively
weak PAH emission. These sources may be intermediate-mass versions of the transitional T Tauri
disks.

\end{itemize}

A larger sample of HAeBe stars will allow us to better determine what a "normal" HAeBe star
looks like. We have begun the next step, a thorough modeling effort including
the addition of detailed PAH spectral synthesis and careful application of
mineralogical models for an even larger sample of stars.

\acknowledgments

We thank the anonymous referee for comments and suggestions that significantly
improved the manuscript. Many thanks to Nirbhik
Chitrakar and Jordan Hyatt for assistance with data analysis and graphics.
L. Keller is grateful to Research Corporation for support of this work. A. Li is supported in part 
by grants from the {\it Spitzer} and HST Theory Programs. Support for this
work was also provided by NASA through contract number 1257184 issued
by JPL/Caltech. This work is based on observations made with the \textit{Spitzer Space
Telescope}, which is operated by the Jet Propulsion Laboratory, California
Institute of Technology under NASA contract 1407. This research has made use of the SIMBAD database
operated at the Centre de Donn\'{e}es astronomiques de Strasbourg
and the NASA Astronomical Data System (ADS).

\clearpage

\begin{deluxetable}{lclccl}
\tablecolumns{6}
\tabletypesize{\small}
\tablewidth{0in}
\tablecaption{Observation Log}

\tablehead{ \colhead{Target\tablenotemark{a}} & \colhead{IRS Campaign} & \colhead{Date} & \colhead{AOR ID}
& \colhead{IRS Module(s)} & \colhead{Comments\tablenotemark{b}}}

\startdata 

HD 32509 & 4 & 2004 Feb.~27 & 3577344 & SL, LL& \\
51 Oph (HD 158643) & 5 & 2004 Mar.~22 & 3582464 & SL, LH & SL saturated \\
HD 152404 (AK Sco) & 20 & 2005 Apr.~14 & 12700160 & SL, LL & \\
HD 141569 & 4 & 2004 Mar.~3 & 3560960 & SL, LH & \\
HD 281789 & 4 & 2004 Feb.~27 & 3529216 & SL, LH & \\
HD 135344 (SAO 206462) & 11 & 2004 Aug.~07 & 3580672 & SL, LH & \\
SU Aur (HD 282624) & 4 & 2004 Feb.~27 & 3533824 & SL, LH & \\
HD 139614 & 11 & 2004 Aug.~07 & 3580928 & SL, LH & \\
HD 142666 & 11 & 2004 Aug.~07 & 3586816 & SL, LH & \\
HD 35187 & 4 & 2004 Feb.~27 & 3578112 & SL, LH & \\
HD 31648 (MWC 480) & 4 & 2004 Feb.~27 & 3577088 & SL, LH & SL saturated \\
HD 145718 & 4 & 2004 Feb.~27 & 3581440 & SL, LH & \\
HD 100453 & 8 & 2004 Jun.~05 & 3578880 & SL, LH & \\
HD 34282 & 13 & 2004 Sep.~26 & 3577856 & SL, LH & \\
HD 144432 & 11 & 2004 Aug.~07 & 3587072 & SL, LH & SL saturated \\
HD 97300 & 29 & 2006 Mar.~06 & 12697088 & SL, LH & \\
HD 169142 & 5 & 2004 Mar.~26 & 3587584 & SL, LH & \\
AB Aur (HD 31293) & 4 & 2004 Feb.~27 & 3533824 & SL, LH & SL saturated \\
Elias 3-1 (V 892 Tau) & 4 & 2004 Feb.~27 & 3529984 & SL, LH & SL saturated \\
HD 97048 & 29 & 2006 Mar.~06 & 12697088 & SL, LH & SL saturated \\
HD 100546 & 30 & 2006 Apr.~06 & 3579136 & SH, LH & No SL \\
HD 104237 (DX Cha) & 21 & 2005 May~05 & 12677632 & SH, LH & No SL\\

\enddata 
\tablenotetext{a}{Alternate target names in parenthesis.}
\tablenotetext{b}{'SL saturated' means the central map position had SL peaking higher than 10 Jy.
In these cases a weighted average of adjacent map positions in SL is substituted. HD 100546 and HD 104237 are both too bright to observe in SL. See text for further details.}

\end{deluxetable}

\begin{deluxetable}{ccc} 
\tablecolumns{3}
\tablewidth{0pt}
\tablecaption{Wavelength intervals used to extract spectral features}
\tablehead{
  \colhead{Feature (\mum)} &
  \colhead{$\lambda_{blue}$ (\mum)} & \colhead{$\lambda_{red}$ (\mum)} }
\startdata
 6.0    &  5.89--5.95  &  6.07--6.13   \\
 6.2    &  5.80--5.95  &  6.54--6.69   \\
 7.7--8.2    &  7.10--7.40  &  8.92--9.17   \\
 8.6    &  8.32--8.44  &  8.86--8.98   \\
11.3    & 10.68--10.86 &  11.82--11.95 \\
12.0    & 11.72--11.90 &  12.13--12.25 \\
12.7    & 12.07--12.19 &  12.98--13.10 \\
\enddata
\end{deluxetable}

\clearpage
\thispagestyle{empty}

\begin{deluxetable}{lccccccc} 
\rotate
\tablecolumns{8}
\tabletypesize{\small}
\tablewidth{0pt}
\tablecaption{Central wavelengths of the PAH features\tablenotemark{1}}
\tablehead{
  \colhead{ } & \multicolumn{6}{c}{$\lambda_{c}$ (\mum)} \\
  \colhead{Target} & \colhead{PAH Classification\tablenotemark{2}} & \colhead{6.2~\mum\ PAH} & \colhead{7.7--8.2~\mum\ PAH} &
  \colhead{8.6~\mum\ PAH} & \colhead{11.3~\mum\ PAH} &
  \colhead{12.0~\mum\ PAH} & \colhead{12.7~\mum\ PAH} }

\startdata

HD 32509  & \nodata & 6.2$\pm$0.1 & 8.2$\pm$0.3 & 8.5$\pm$0.2\tablenotemark{a}
                                                                & 11.3$\pm$0.1 & \nodata          & 12.3$\pm$0.1\tablenotemark{b}\\
51 Oph    & C & \nodata         & 8.5$\pm$0.1 & \nodata         & \nodata          & \nodata          & \nodata         \\
HD 152404 & C & 6.27$\pm$0.03 & 8.34$\pm$0.05\tablenotemark{a}
                                              & 8.67$\pm$0.05 & 11.27$\pm$0.03 & \nodata          & 12.7$\pm$0.2\\
HD 141569 & B/C & 6.26$\pm$0.01 & 7.96$\pm$0.01 & 8.67$\pm$0.03 & 11.28$\pm$0.02 & \nodata          & 12.49$\pm$0.07\\
HD 281789 & B/C & 6.25$\pm$0.01 & 7.92$\pm$0.02 & 8.65$\pm$0.02 & 11.25$\pm$0.04 & \nodata          & 12.64$\pm$0.07\\
HD 135344 & B/C & 6.28$\pm$0.02 & 8.05$\pm$0.04 & 8.64$\pm$0.07 & 11.34$\pm$0.01 & 12.07$\pm$0.07\tablenotemark{b}
                                                                                                      & 12.69$\pm$0.06\\
SU Aur  &  C & 6.27$\pm$0.01 & 8.18$\pm$0.01 & \nodata         & 11.32$\pm$0.01 & \nodata          & 12.74$\pm$0.08\\
HD 139614 & B/C & 6.277$\pm$0.004 & 8.04$\pm$0.01 & 8.68$\pm$0.02 & 11.32$\pm$0.01 & \nodata          & 12.66$\pm$0.11\\
HD 142666 & B/C & 6.25$\pm$0.01 & 8.06$\pm$0.03 & 8.65$\pm$0.03 & 11.28$\pm$0.06 & \nodata          & 12.65$\pm$0.03\\
HD 35187  & B/C & 6.27$\pm$0.01 & 8.04$\pm$0.01 & 8.66$\pm$0.02 & 11.30$\pm$0.04 & \nodata          & 12.66$\pm$0.05\\
HD 31648  & C & 6.28$\pm$0.01 & 8.3$\pm$0.1 & \nodata         & 11.25$\pm$0.05 & \nodata          & 12.5$\pm$0.1\\
HD 145718 & B/C & 6.258$\pm$0.004 & 8.05$\pm$0.02 & 8.65$\pm$0.03 & 11.28$\pm$0.01 & 12.0$\pm$0.1\tablenotemark{a}  & 12.63$\pm$0.06\\
HD 100453 & B/C & 6.28$\pm$0.01 & 8.08$\pm$0.01 & 8.70$\pm$0.03 & 11.33$\pm$0.05 & 12.05$\pm$0.08 & 12.74$\pm$0.06\\
HD 34282  & B/C & 6.261$\pm$0.003 & 7.93$\pm$0.02 & 8.65$\pm$0.01 & 11.27$\pm$0.01 & 11.96$\pm$0.08\tablenotemark{a}  & 12.63$\pm$0.02\\                                                                                                    
HD 144432 & C & 6.27$\pm$0.02 & 8.39$\pm$0.05 & 8.66$\pm$0.07 & 11.26$\pm$0.03 & \nodata          & 12.59$\pm$0.07\\
HD 97300  & A & 6.233$\pm$0.004 & 7.81$\pm$0.04 & 8.63$\pm$0.02 & 11.25$\pm$0.01 & 11.98$\pm$0.03\tablenotemark{b}  & 12.64$\pm$0.01\\
HD 169142 & B/C & 6.261$\pm$0.003 & 7.99$\pm$0.01 & 8.67$\pm$0.01 & 11.30$\pm$0.01 & 12.03$\pm$0.02 & 12.70$\pm$0.02\\
AB Aur    & B/C & 6.25$\pm$0.01 & 7.95$\pm$0.04 & 8.64$\pm$0.07 & 11.26$\pm$0.03 & \nodata          & 12.50$\pm$0.09\\
Elias 3-1 & B/C & 6.245$\pm$0.002 & 7.81$\pm$0.04 & 8.64$\pm$0.03 & 11.23$\pm$0.06 & \nodata          & 12.63$\pm$0.09\\
HD 97048  & B/C & 6.234$\pm$0.002 & 7.87$\pm$0.03 & 8.63$\pm$0.01 & 11.26$\pm$0.01 & 12.00$\pm$0.04 & 12.68$\pm$0.01\\
HD 100546 & B/C & no SL         & no SL        & no SL  & 11.23$\pm$0.01 & 11.94$\pm$0.03 & 12.63$\pm$0.05\\
HD 104237 & B/C & no SL         & no SL        & no SL  & 11.18$\pm$0.03 & 12.0$\pm$0.1\tablenotemark{a}
                                                                                                      & 12.4$\pm$0.3\tablenotemark{a}\\

\enddata
\tablenotetext{1}{Missing data indicated by "\nodata" are features for which the S/N in the spectrum was $<$1.}
\tablenotetext{2}{PAH spectral classification according to \cite{pee02}.}
\tablenotetext{a}{S/N of extracted feature $<$ 2.}
\tablenotetext{b}{S/N of extracted feature $<$ 3.}
\end{deluxetable}

\clearpage

\begin{deluxetable}{lrrrrrr} 
\tablecolumns{7}
\tablewidth{0pt}
\tabletypesize{\small}
\tablecaption{Extracted strengths of the PAH features}
\tablehead{
  \colhead{ } & \multicolumn{6}{c}{$F$ (10$^{-15}$ W/m$^2$)} \\
  \colhead{Target} & \colhead{6.2~\mum} & \colhead{7.7--8.2~\mum} &
  \colhead{8.6~~\mum} & \colhead{11.3~\mum} & \colhead{12.0~\mum} &
  \colhead{12.7~\mum}}
\startdata

HD 32509  &    0.5$\pm$0.1 &   0.4$\pm$0.1 &  0.1$\pm$0.1 &   0.3$\pm$0.1 & $<$0.01 &    0.09$\pm$0.03\\
51 Oph    & $<$1 &   9$\pm$3 &  $<$2 &   $<$8 &    $<$0.3 & $<$0.7\\
HD 152404 &    1.1$\pm$0.1 &   4$\pm$3 &  0.5$\pm$0.1 &   2.5$\pm$0.1 &    $<$0.03 &    0.2$\pm$0.1\\
HD 141569 &    8.7$\pm$0.4 &  19.2$\pm$0.4 &  1.1$\pm$0.1 &   0.7$\pm$0.1 &    $<$0.1 &    0.33$\pm$0.04\\
HD 281789 &    4.3$\pm$0.1 &   7.8$\pm$0.2 &  1.1$\pm$0.1 &   1.8$\pm$0.2 &    $<$0.1 &    0.4$\pm$0.1\\
HD 135344 &    5.2$\pm$0.6 &  11.6$\pm$0.6 &  0.4$\pm$0.1 &   2.5$\pm$0.1 &    0.05$\pm$0.02 &    0.4$\pm$0.1\\
SU Aur    &    5.9$\pm$0.3 &   8.4$\pm$0.1 &  $<$0.1 &   4.3$\pm$0.1 &    $<$0.4 &    0.4$\pm$0.1\\
HD 139614 &    7.0$\pm$0.2 &  17.1$\pm$0.3 &  1.4$\pm$0.1 &   3.9$\pm$0.1 &    $<$0.1 &    0.6$\pm$0.1\\
HD 142666 &   10.9$\pm$0.3 &  16.1$\pm$0.4 &  1.8$\pm$0.2 &   5.3$\pm$0.8 &    $<$0.1 &    1.0$\pm$0.1\\
HD 35187  &    8.3$\pm$0.3 &  20.6$\pm$0.2 &  1.8$\pm$0.1 &   3.4$\pm$0.4 &    $<$0.2 &    0.6$\pm$0.1\\
HD 31648  &    9.5$\pm$0.7 &  21$\pm$4 &  $<$3 &  16$\pm$2 & $<$0.2 &    2.1$\pm$0.4\\
HD 145718 &   16.1$\pm$0.3 &  27.2$\pm$0.5 &  2.5$\pm$0.3 &   9.1$\pm$0.2 &    0.1$\pm$0.1 &    1.0$\pm$0.1\\
HD 100453 &   26$\pm$1 &  37$\pm$1 &  2.8$\pm$0.3 &  12$\pm$2 &    0.24$\pm$0.01 &    1.5$\pm$0.3\\
HD 34282  &   12.4$\pm$0.2 &  39$\pm$1 &  2.6$\pm$0.1 &   3.8$\pm$0.1 &    0.02$\pm$0.01 &    0.66$\pm$0.03\\
HD 144432 &    2.2$\pm$0.2 &   10$\pm$1 &  0.9$\pm$0.3 &  11$\pm$1 &    $<$0.1 &    0.7$\pm$0.1\\
HD 97300  &   13.4$\pm$0.2 &  40$\pm$2 &  4.5$\pm$0.3 &   7.6$\pm$0.2 &    0.2$\pm$0.1 &    2.22$\pm$0.04\\
HD 169142 &   31.3$\pm$0.5 &  74$\pm$1 &  5.2$\pm$0.2 &  12.8$\pm$0.2 &    0.23$\pm$0.03 &    2.1$\pm$0.1\\
AB Aur    &   66$\pm$2 & 125$\pm$5 & 12$\pm$3 &  23$\pm$2 & $<$0.3 &    5$\pm$1\\
Elias 3-1 &   58$\pm$1 & 112$\pm$9 & 24$\pm$3 &  25$\pm$4 &    $<$2 &    5$\pm$1\\
HD 97048  &   58$\pm$1 & 196$\pm$8 & 19$\pm$1 &  45$\pm$1 &    0.6$\pm$0.1 &   11.5$\pm$0.2\\
HD 100546  &  no SL         & no SL         & no SL        & 132$\pm$6 &    5$\pm$2 &   12$\pm$2\\
HD 104237 &  no SL         & no SL        & no SL        &  16$\pm$1 &    0.7$\pm$0.5 &    1.8$\pm$0.9\\

\enddata

\end{deluxetable}

\begin{deluxetable}{lclcccccl} 
  \rotate \tablecolumns{9} \tablewidth{0in} \tabletypesize{\small}
  \tablecaption{Stellar Physical Characteristics} \tablehead{
    \colhead{Star Name} & \colhead{$T_{e}(K)$} & \colhead{$Sp_{type}$}
    & \colhead{$m_V$} & \colhead{d (pc)} & \colhead{SED
      Group\tablenotemark{a}} & \colhead{$A_V$\tablenotemark{b}} &
    \colhead{$i$ (deg)} & \colhead{Sp. Type Ref.}}
 
\startdata
HD 32509 & 8970 & A2e & 7.51 &  151 & IIb\tablenotemark{1} & \nodata & 51 & \cite{mal98} \\
51 Oph & 9520 & A0Ve & 4.78  & 131 & IIa & 0.11 & 35 & \cite{mal98} \\
HD 152404 & 6440 & F5 Ve & 9.14 & 145 & IIa & 0.62 & \nodata & \cite{mal98} \\ 
HD 141569 & 9520 & A0 Ve & 7.09 & 99 & IIb & 0.37 & 51 & \cite{dmr97} \\ 
HD 281789 & 9520 & A0 & 10.01 & 350 & IIa\tablenotemark{1} & \nodata & \nodata & \cite{nes95} \\
HD 135344 & 6590 & F4 Ve & 8.61 & 140 & Ib & 0.31 & 45 & \cite{dmr97} \\
SU Aur & 5945 & G1 & 9.42 &  152 & IIa\tablenotemark{1} & \nodata & \nodata & \cite{cal04} \\
HD 139614 & 7850 & A7 Ve & 8.26 & 140 & Ia & 0.09 & 20 & \cite{hou78} \\
HD 142666 & 7580 & A8 Ve & 8.81 & 145 & IIa & 0.93 & 55 & \cite{hou88} \\
HD 35187 & 8970 & A2 Ve & 7.78 &  150 & IIa & 0.71 & \nodata & \cite{dun98} \\
HD 31648 & 8970 & A2 pshe & 7.72  & 131 & IIa & 0.25 & 38 & \cite{mal98}\\
HD 145718 & 7580 & A8 III/IVe & 8.98 &  131 & IIa\tablenotemark{1} & \nodata &  \nodata & \cite{hou88} \\
HD 100453 & 7390 & A9 Ve & 7.79 & 112 & Ib & 0.02 & \nodata & \cite{hou75} \\
HD 34282 & 9333 & A0.5 Vbe & 10.11 & 160 & Ib & 0.28 & 56 & \cite{gra98}\\
HD 144432 & 7390 & A9 Ve & 8.15 & 253 & IIa & 0.17 & 45 & \cite{mal98} \\
HD 97300 & 10500 & B9 V & 9.00 &  188 & IIb\tablenotemark{1} & \nodata & \nodata & \cite{hou75} \\
HD 169142 & 8200 & A5 Ve & 8.11  & 145 & Ib & 0.43 & 8 & \cite{mal98} \\
AB Aur & 9520 & A0 pe & 7.05 & 144 & Ia & 0.50 & 65 & \cite{rac68} \\
Elias 3-1 &  10500 & B9 & 15.3 & 160 &  Ia & 4.05 & $<$ 45 &  \cite{str94} \\
HD 97048 & 9520 & A0 pshe & 8.46 &175 & Ib & 1.26 & \nodata & \cite{hou75} \\
HD 100546 & 10500 & B9 Vne & 6.70 & 103 & Ia & 0.26 & \nodata & \cite{mal98} \\
HD 104237 & 7580 & A8 pe & 6.81 & 116 & IIa & 0.29 & 18 & \cite{mal98} \\

\enddata 

\tablenotetext{a}{SED Group classifications after \cite{mee01} and quoted from \cite{aa04}.}
\tablenotetext{b}{Visual extinction values quoted from \cite{aa04}, except for 51 Oph \citep{mal98}.}
\tablenotetext{1}{SED Group for SU Aur, HD 281789, HD 32509, HD 97300 determined using the criterion of \cite{mee01}. }
\tablecomments{V892 Tau: distance assumed to be 160 pc (Taurus) \\
 T$_e$from Kenyon and Hartman (1995) Table A5 for all luminosity class
V, other luminosity class data from C. W. Allen (2000, Astrophysical Quantities, ed. A. N. Cox, Springer)}

\end{deluxetable}

\clearpage

%
\begin{figure}
\includegraphics[width=6in]{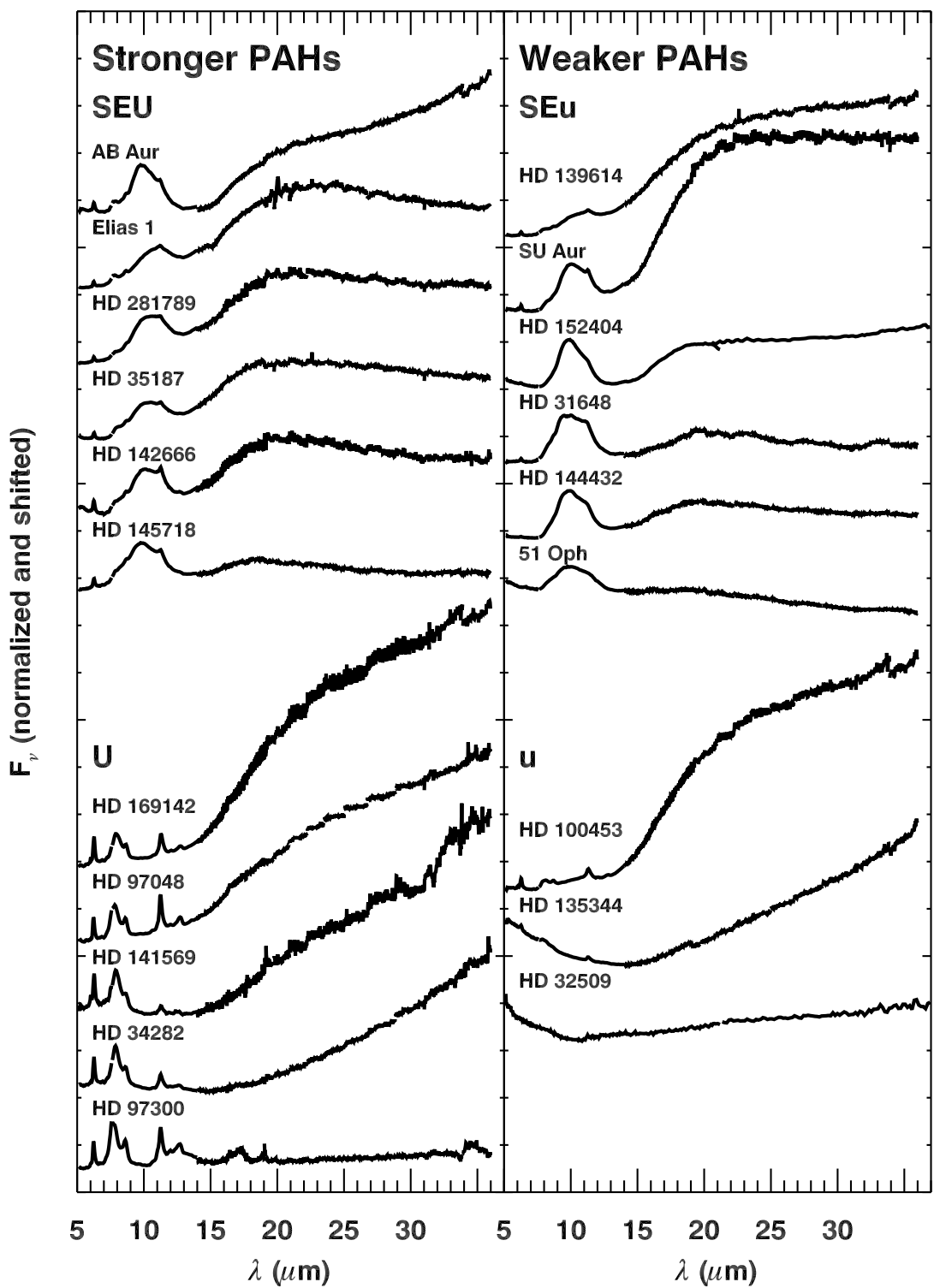} 
\caption{Spectra of all of our program stars plotted in
$\lambda$  vs. F$_{\nu}$ units, but arbitrarily scaled to fit on the same axes for easy comparison of the general shape
of the SEDs.  See Figure 2 for individual, flux-calibrated spectra. The designations 'U' (Stonger PAHs)
and 'u' (Weaker PAHs) refer to the contrast of the PAH features relative to the continuum.}
\end{figure}

\clearpage

%
\begin{figure}
\includegraphics[width=5in]{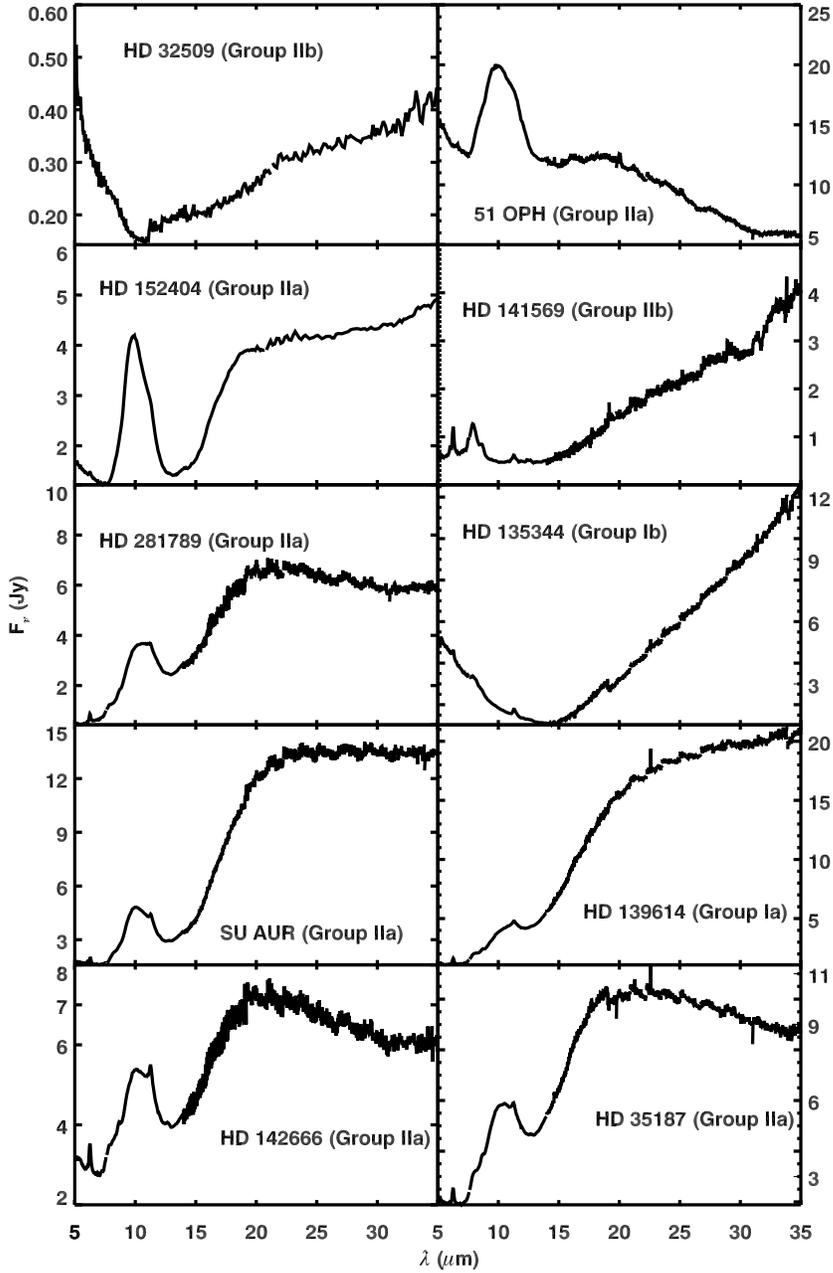}
\caption{All spectra flux calibrated and displayed on individual
axes in $\lambda$ vs. $F_{\nu}$ units. Spectra are labeled with star name and Meeus et all. (2001)
SED group classification (I, Ia/b, II, IIa/b).}
\end{figure}
\clearpage
\begin{center}
\includegraphics[width=5in]{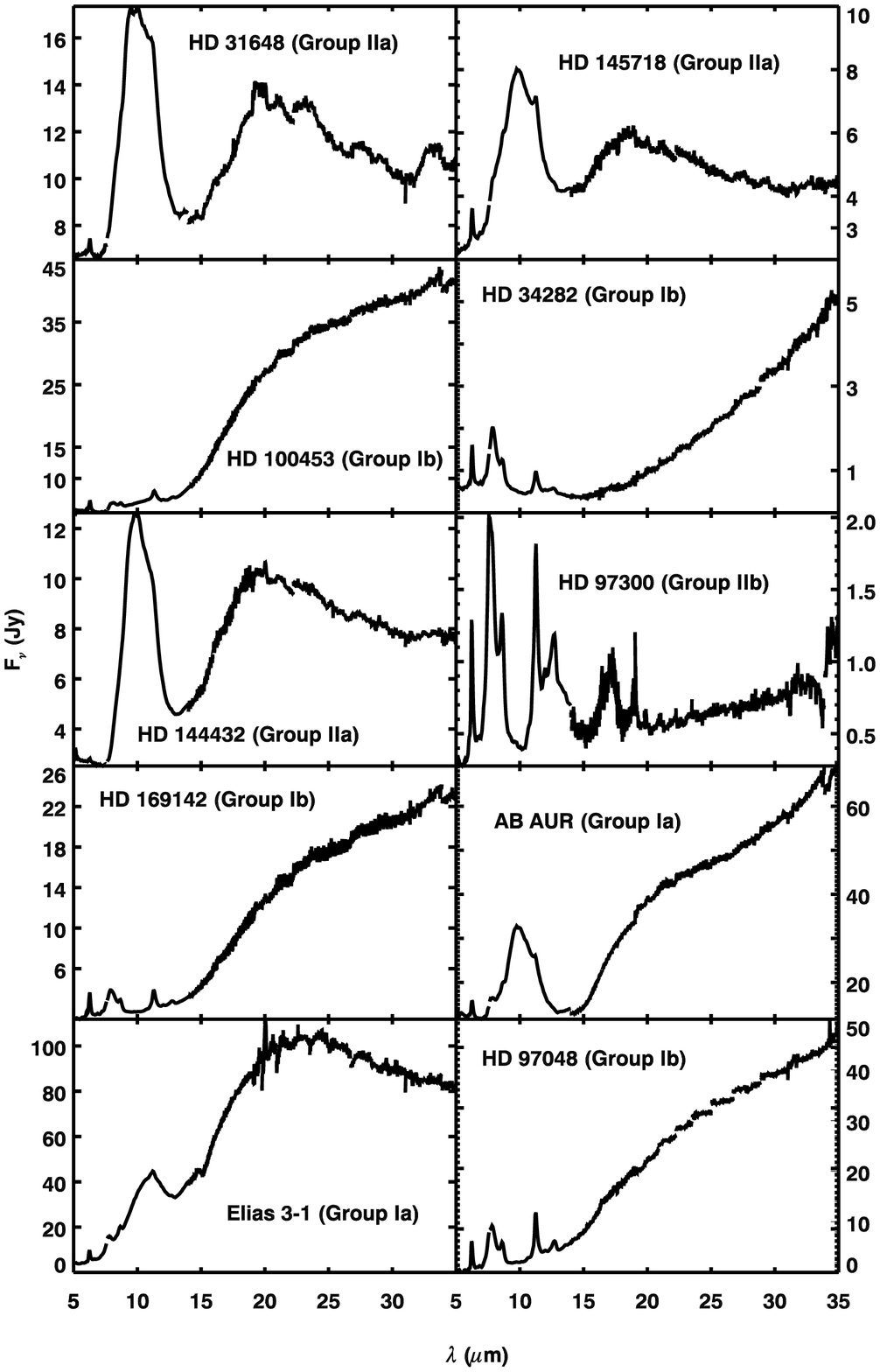}\\
{Fig. 2. --- Continued.}
\end{center}
\clearpage
\begin{center}
\includegraphics[width=5in]{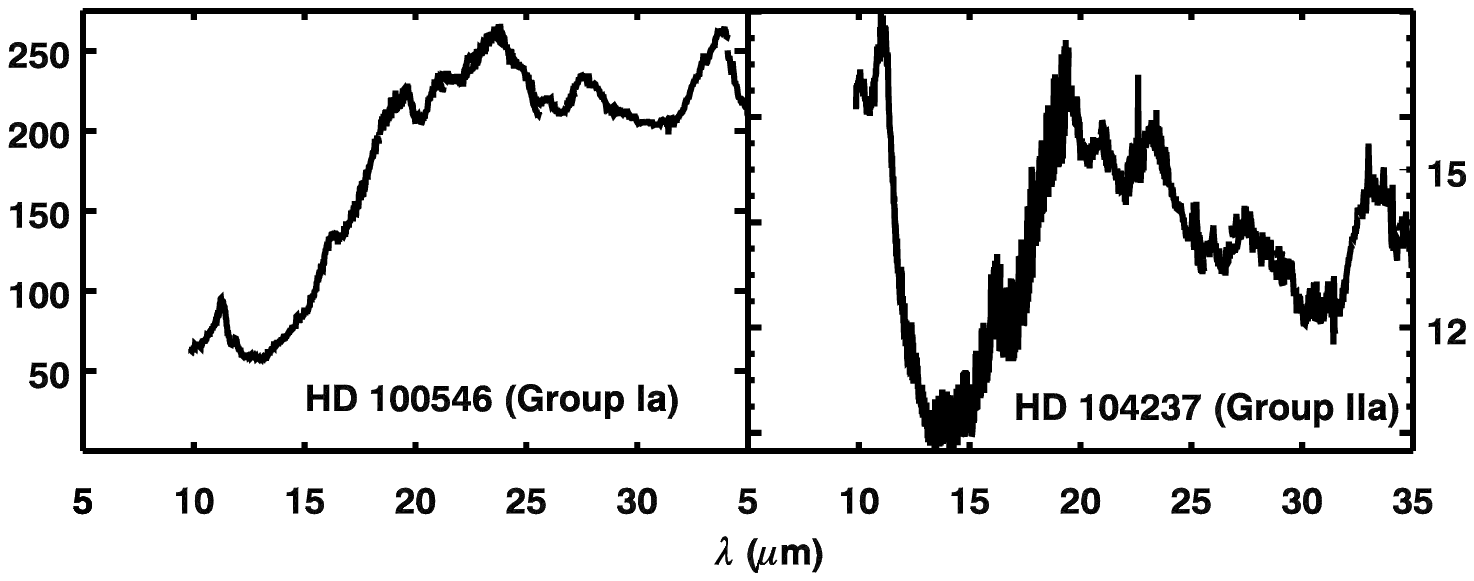}\\
{Fig. 2. --- Continued.}
\end{center}

\clearpage

%
\begin{figure}
\includegraphics[width=5in]{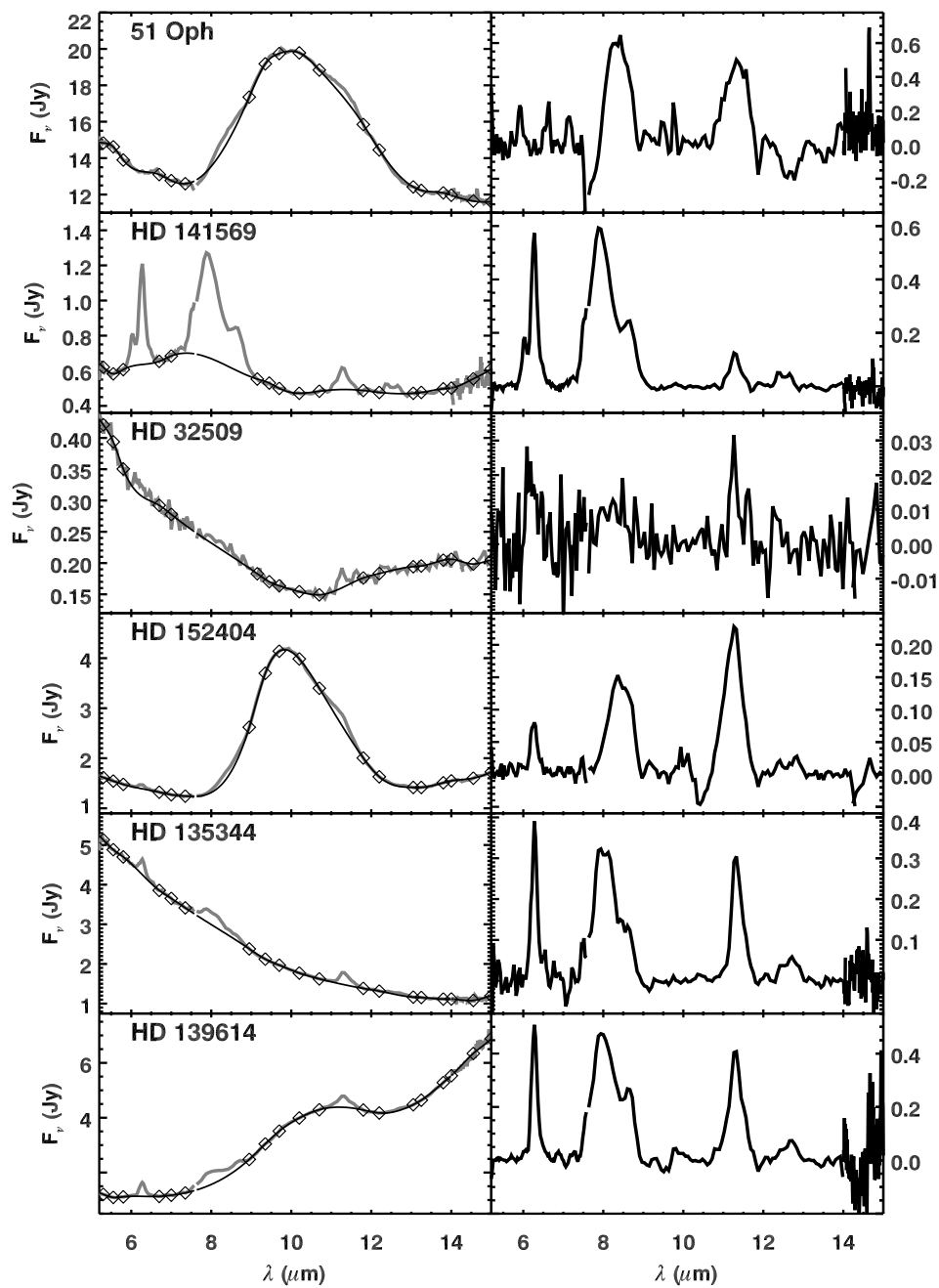}
\caption{Several examples of  spectra with dust continuum and dust features fit with spline
curves (left, with anchor wavelengths
of the spline fits indicated by open diamonds) and 
continuum-subtracted (right), leaving the PAH spectra for
our analysis. Examples range from 'weak or no PAH' (e.g. 51 Oph and HD 32509) to 'strong PAH'.}
\end{figure}

\clearpage

%
\begin{figure}
\includegraphics[width=6in]{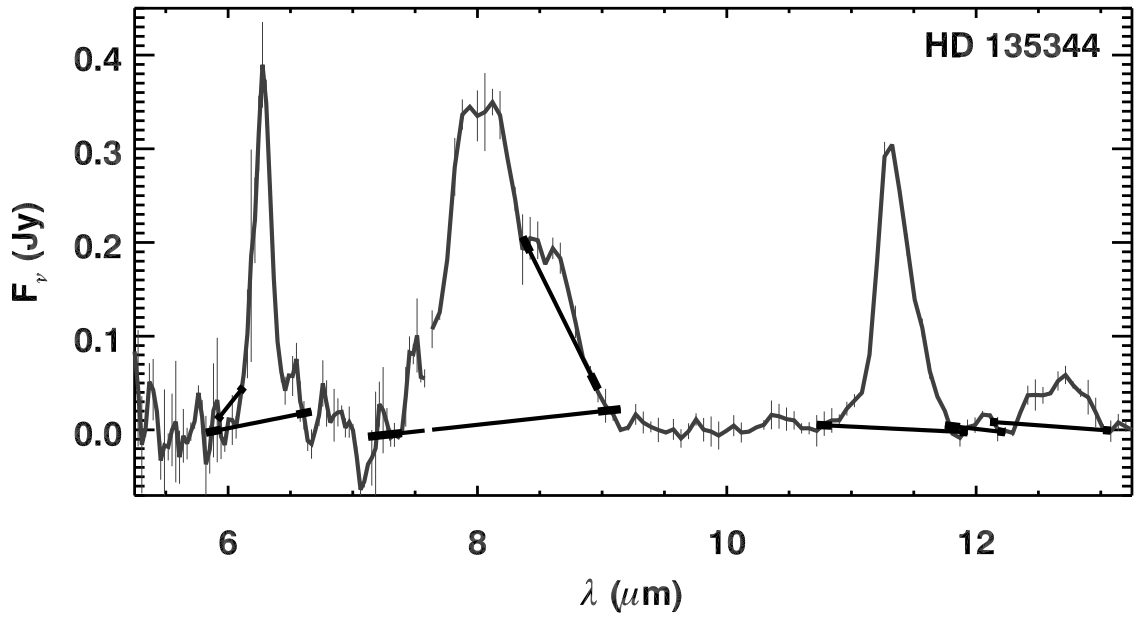}
\caption{The continuum-subtracted spectrum of HD 135344, a relatively weak PAH emitter
among our sample stars, and the linear fits used to define center wavelengths and extract the integrated fluxes
of the PAH features after the spline fit is subtracted. We present the wavelengths used for this extraction
method in Table 2.}
\end{figure}

\clearpage

%
\begin{figure}
\includegraphics[width=6in]{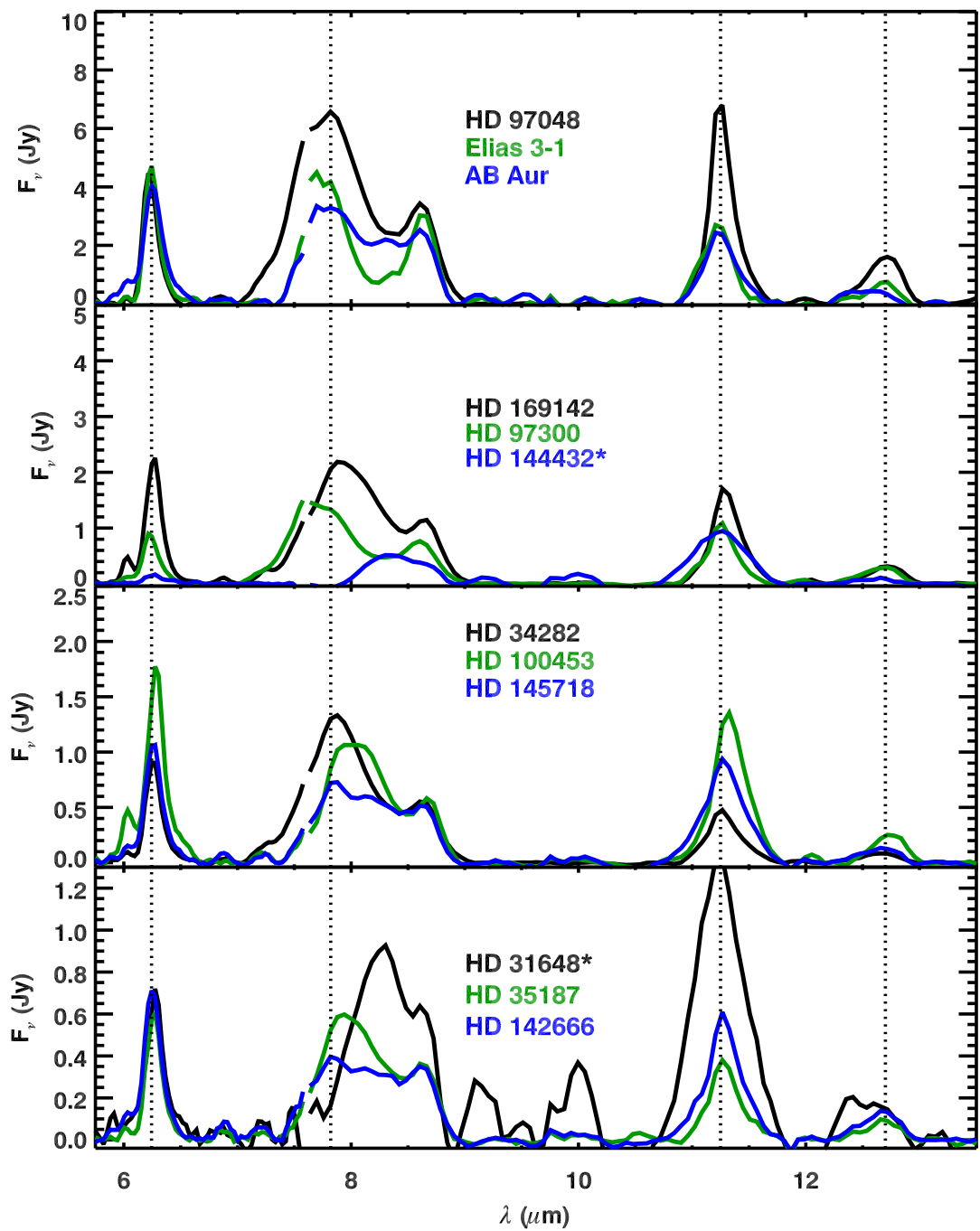} 
\caption{Continuum-subtracted PAH spectra for all sources. Dotted vertical lines indicate the
nominal wavelengths for the 6.2, 11.3, and 12.7~\mum\ features as well as for the nominal 
peak of the 7.7--8.2~\mum\ complex. Sources arranged roughly in order of decreasing PAH
integrated flux (top to bottom). Asterisks next to star names indicate sources that are not
included in Figures 6-9 because their PAH feature wavelengths are uncertain.}
\end{figure}
\clearpage
\begin{center}
\includegraphics[width=6in]{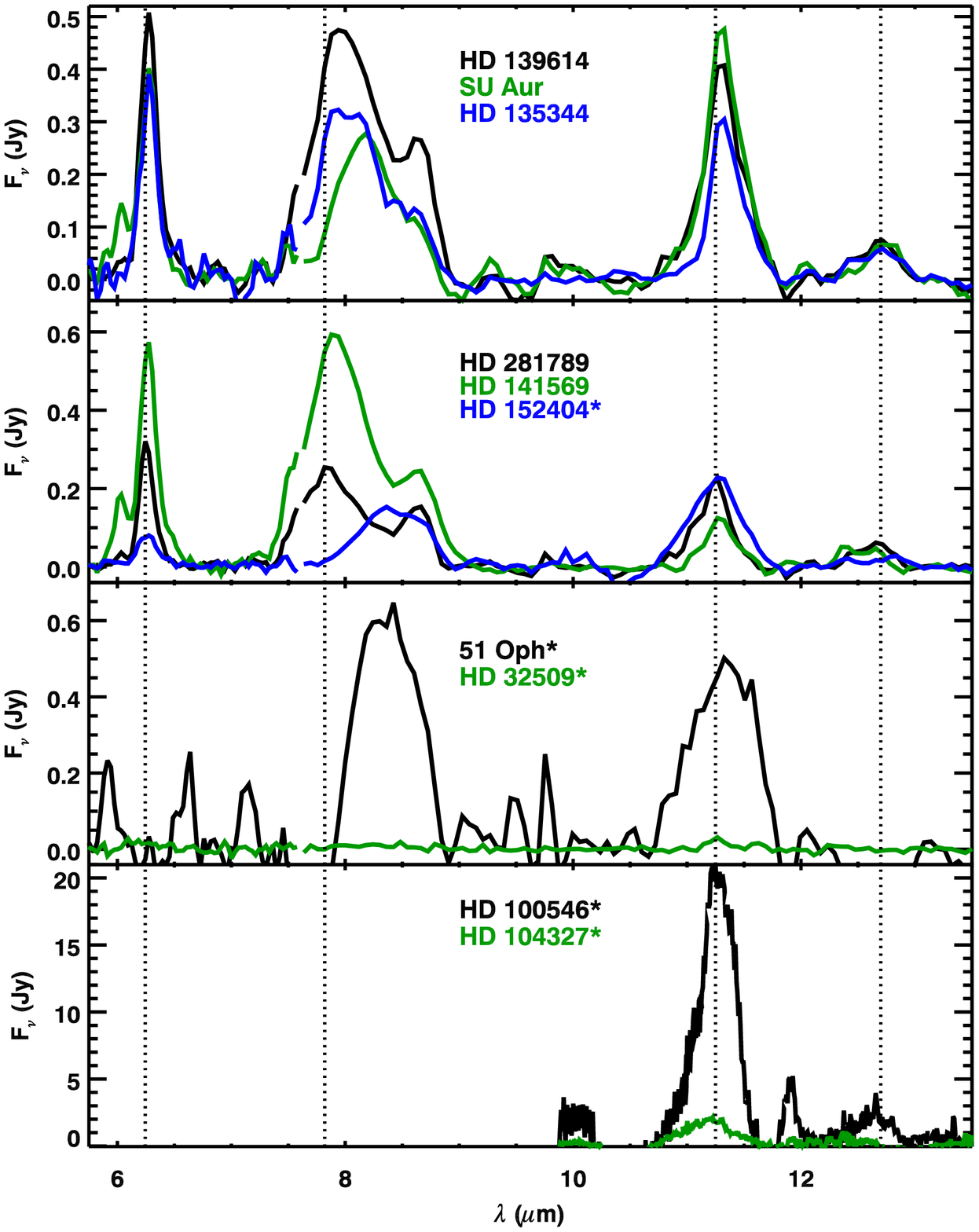}\\
{Fig. 5. --- Continued.}
\end{center}

\clearpage

%
\begin{figure}
\includegraphics[width=5in]{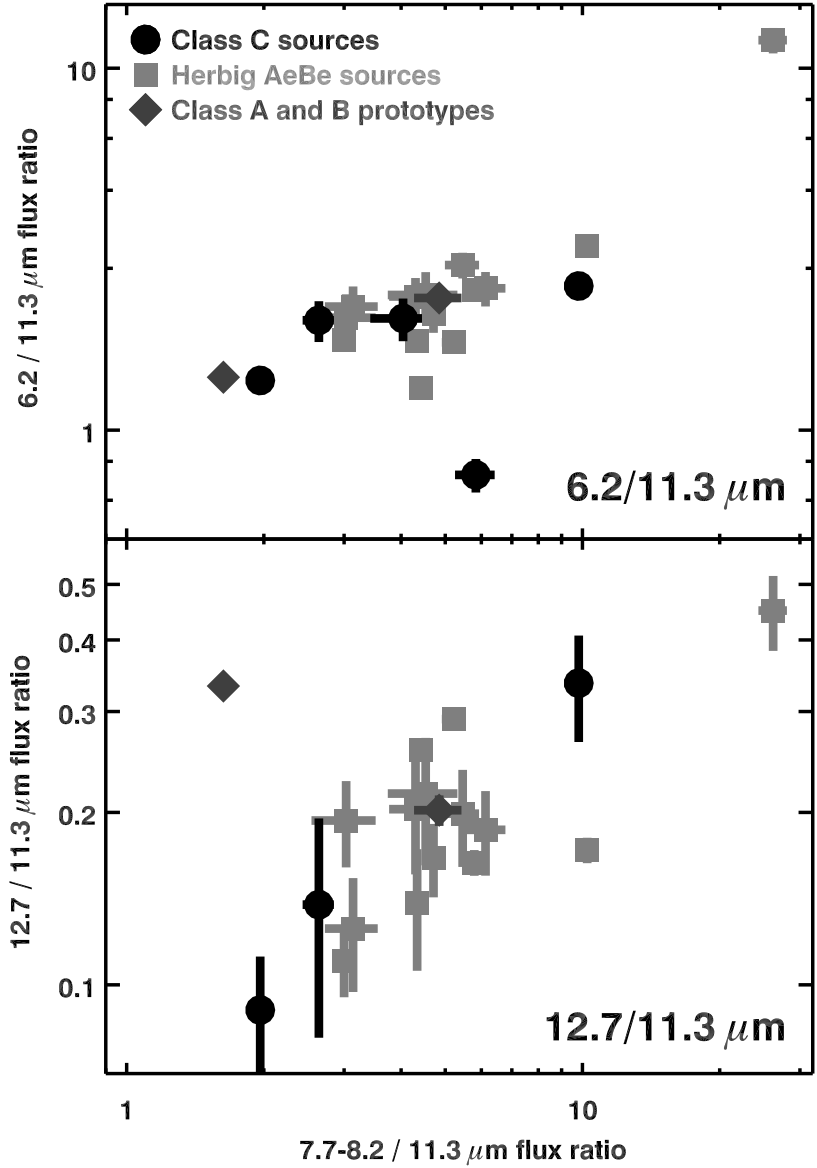}
\caption{Flux ratios from the different PAH features plotted vs. the ratio
of the 7.9 and 11.3~\mum\ PAH features on a log-log scale. The ratio $F_{7.7-8.2}/F_{11.3}$
traces the ionization fraction of the PAHs. The most ionized PAH spectrum is
from HD 141569 (upper right data point in both plots). In the top panel, $F_{6.2}/F_{11.3}$
increases with ionization of the PAHs (Pearson correlation coefficient, r=0.94). $F_{12.7}/F_{11.3}$ also increases with ionization fraction, suggesting that the size of the PAHs decreases with ionization fraction (correlation coefficient, r=0.69). The Class A and B prototypes and the Class C sources are from \cite{slo07} and
we list them in Section 2.5. The A, B, C spectral classification system is described in Section 2 of the text.}
\end{figure}

\clearpage

%
\begin{figure}
\includegraphics[width=6in]{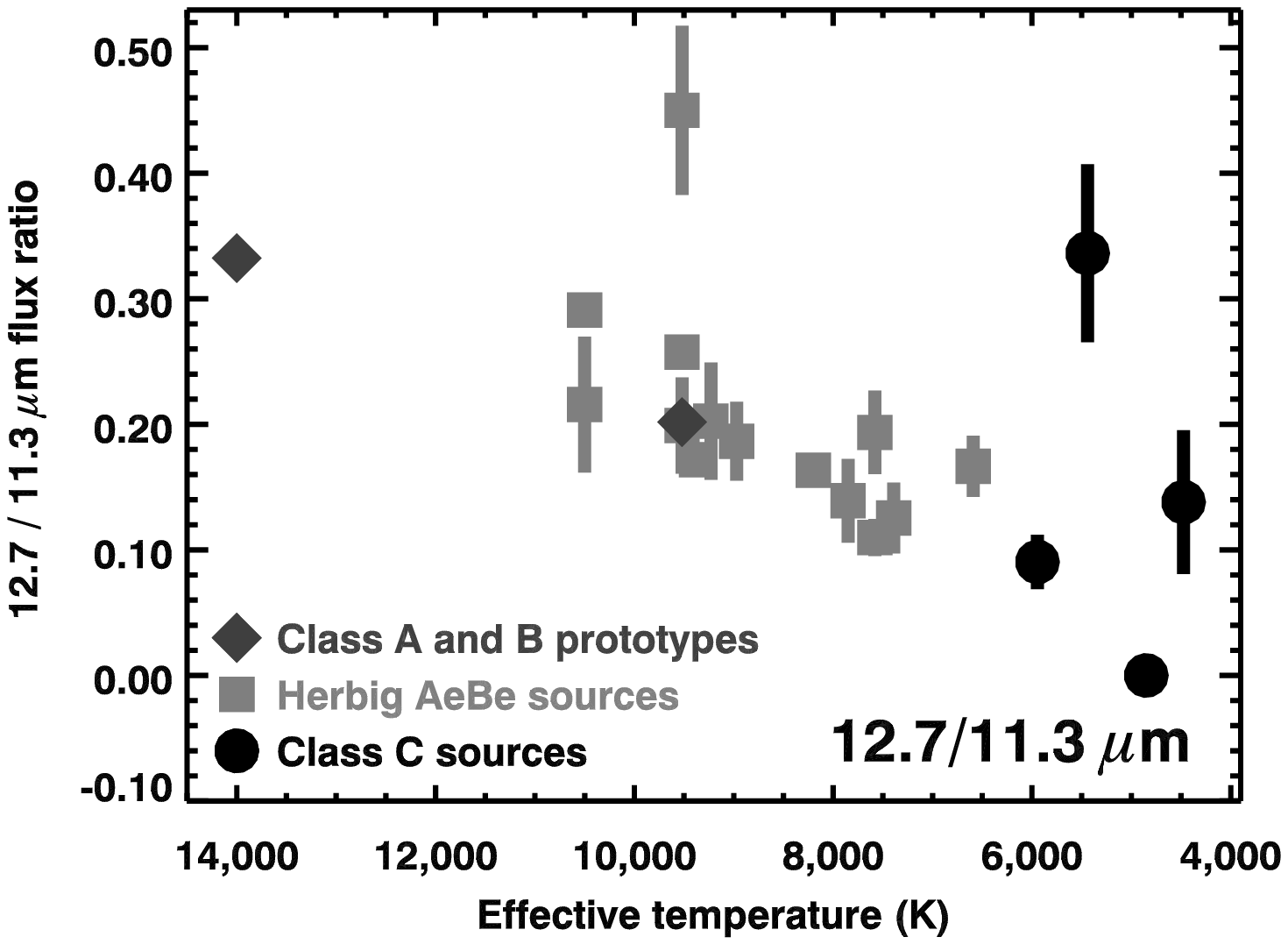} 
\caption{PAH $F_{12.7}/F_{11.3}$ vs. effective temperature. If the ratio of the 
solo C-H bend (11.3~\mum) to the trio C-H bend (12.7~\mum) increases when PAH 
molecules size is smaller, then these data indicate that PAH molecule size decreases with
increasing stellar effective temperature (Pearson correlation coefficient, r=0.85). Class A, B, 
and C prototypes are included to more
clearly illustrate that HAeBe stars clump in the Class B-C range. The outlying source at
top center is HD 141569.}
\end{figure}

\clearpage

%

\begin{figure}
\includegraphics[width=6in]{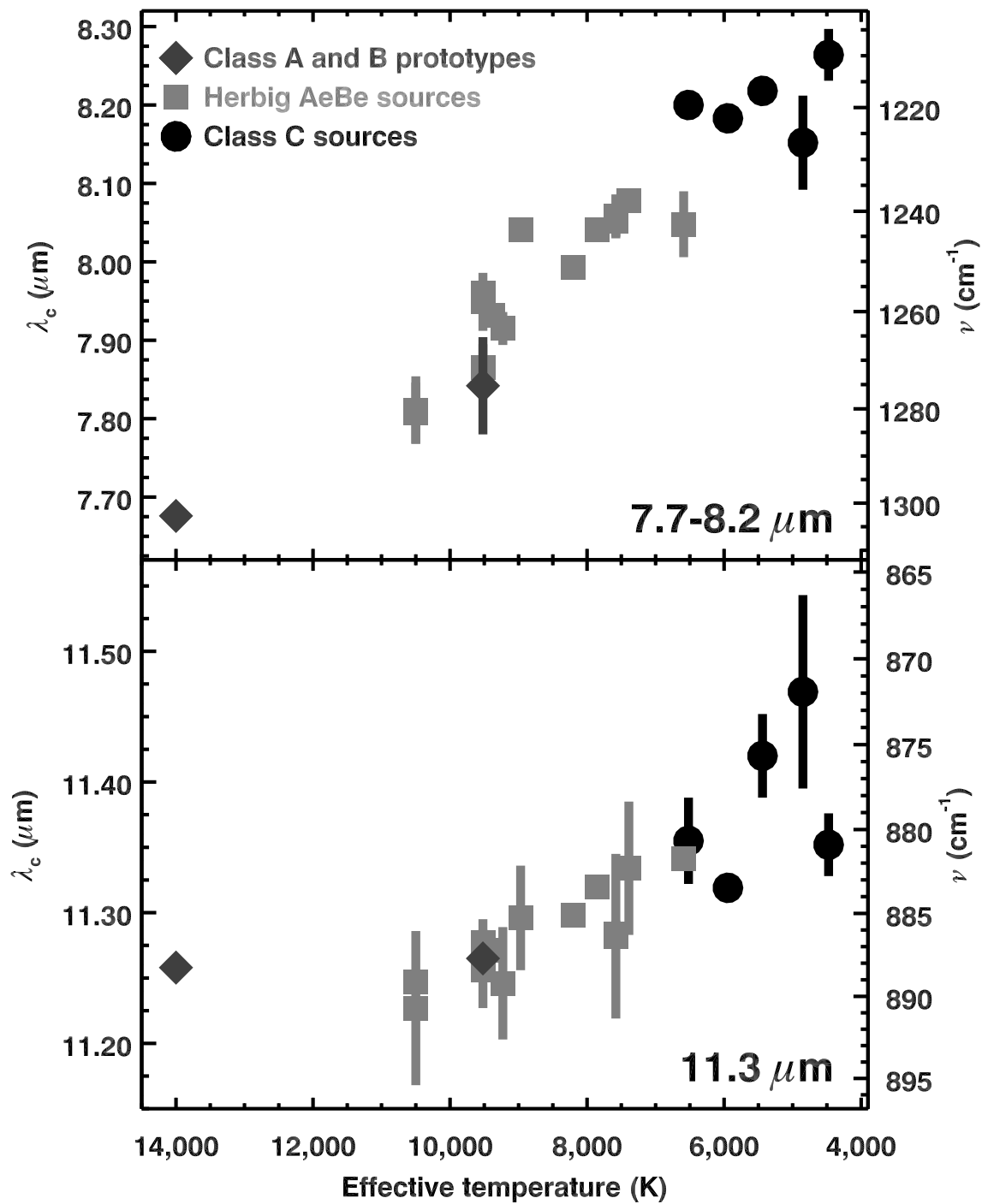}
\caption{Wavelengths of the 7.7--8.2~\mum\ and 11.3~\mum\ PAH features plotted vs. effective temperature
of the host star. Pearson correlation coefficients are r=0.95 and r=0.79 for the 7.7--8.2~\mum\ and 11.3~\mum\ features
respectively.} 
\end{figure}

\clearpage

%

\begin{figure}
\includegraphics[width=6in]{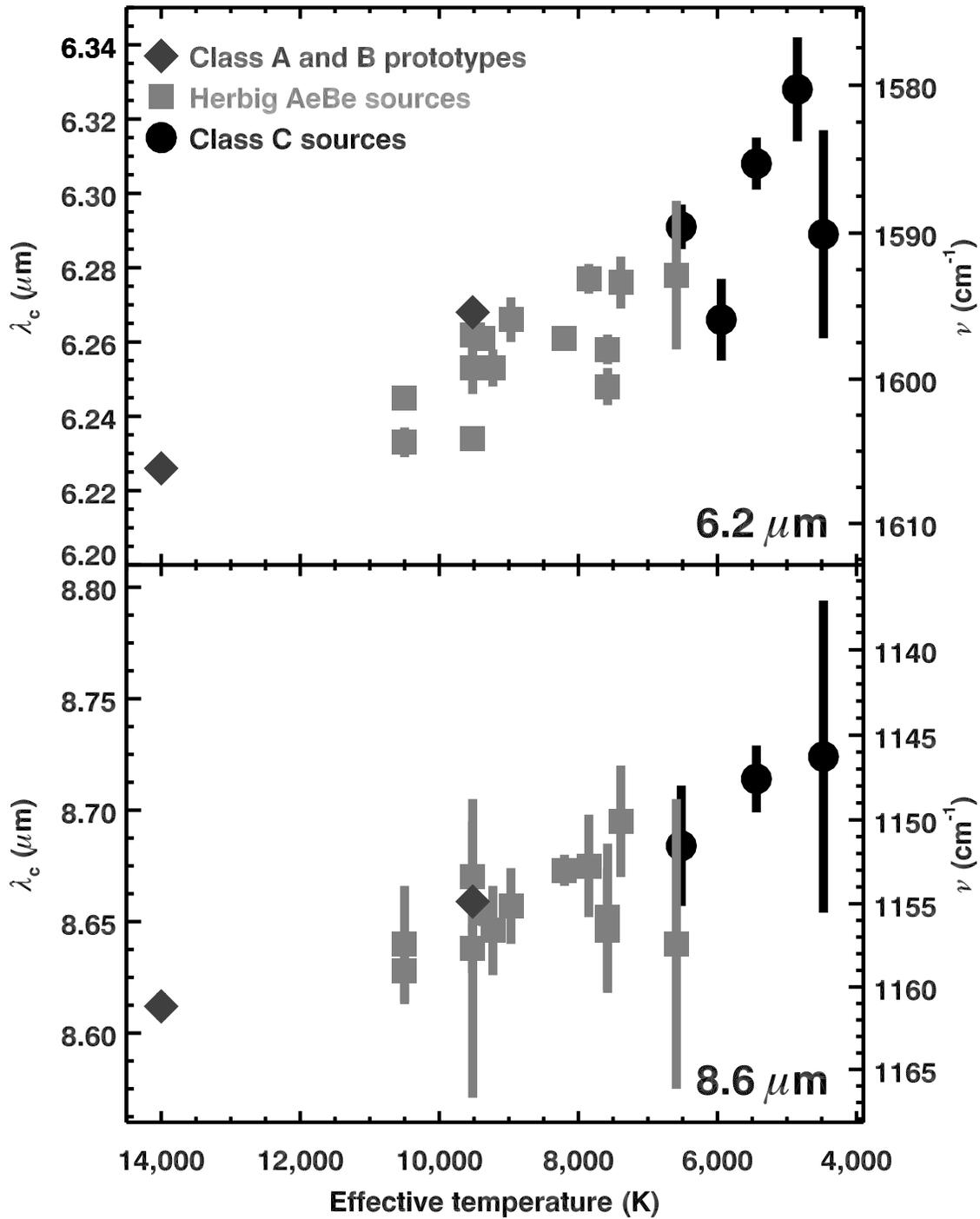}
\caption{Wavelengths of the 6.2~\mum\ and 8.6~\mum\ PAH features plotted vs. effective temperature
of the host star. Pearson correlation coefficients are r=0.83 and r=0.81 for the 6.2~\mum\ and 8.6~\mum\ features
respectively.} 
\end{figure}

\clearpage

%

\begin{figure}
\includegraphics[width=6in]{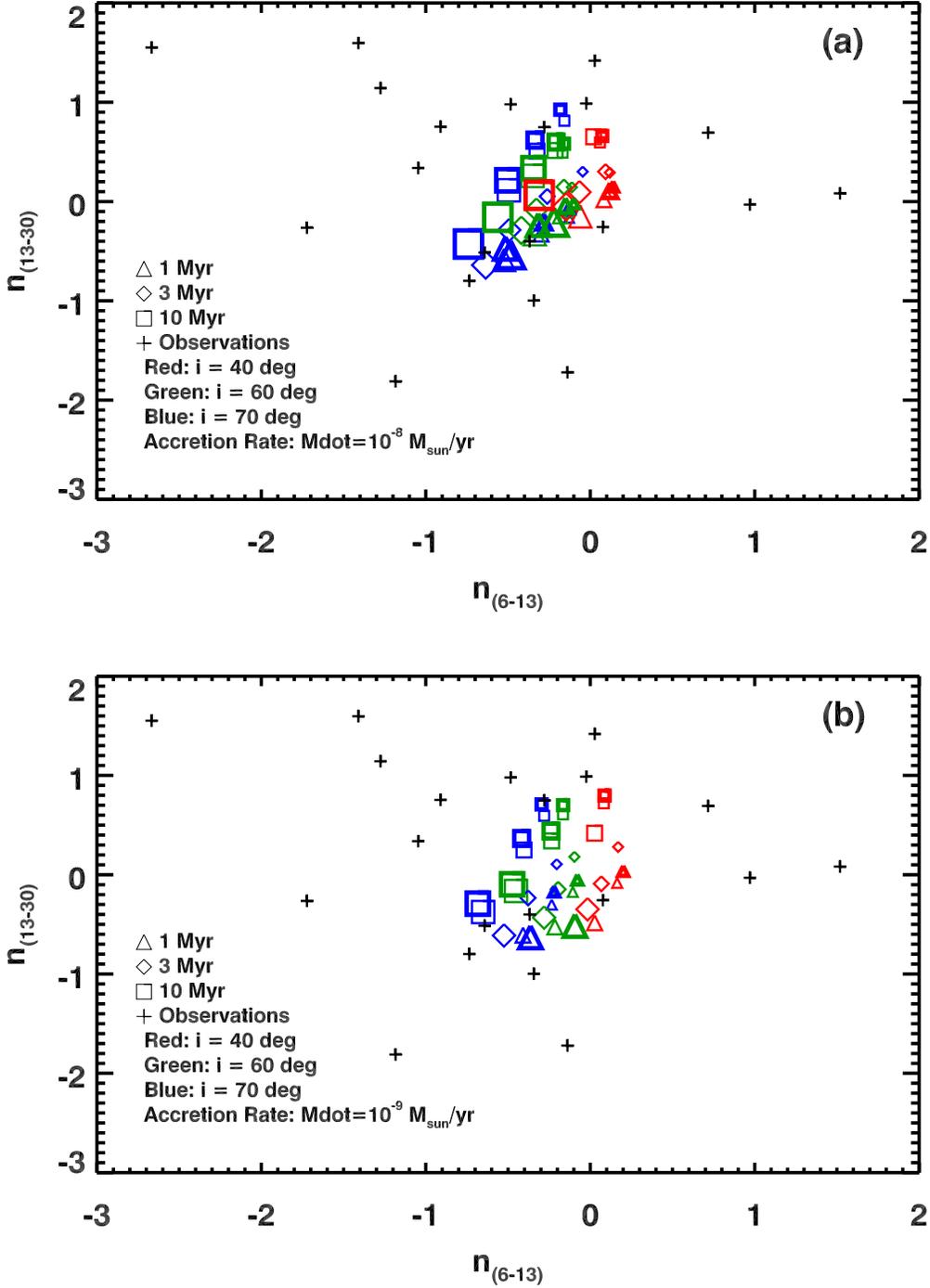}
\caption{Continuum spectral indices obtained from model spectral energy distributions: (a) with 
dust settling in the inner wall of the disk and accretion rates $\dot{M}=10^{-8} M_\odot/yr$; (b) with dust settling in the inner wall and  $\dot{M}=10^{-9} M_\odot/yr$; (c) with NO dust settling in the inner wall and $\dot{M}=10^{-8} M_\odot/yr$; (d) with NO dust settling in the inner wall and $\dot{M}=10^{-9} M_\odot/yr$. Symbol sizes increase with dust settling (e.g. increasing $\epsilon$), thin symbols are for spectral type A2, bold symbols are for spectral type A6.} 
\end{figure}
\clearpage
\begin{center}
\includegraphics[width=6in]{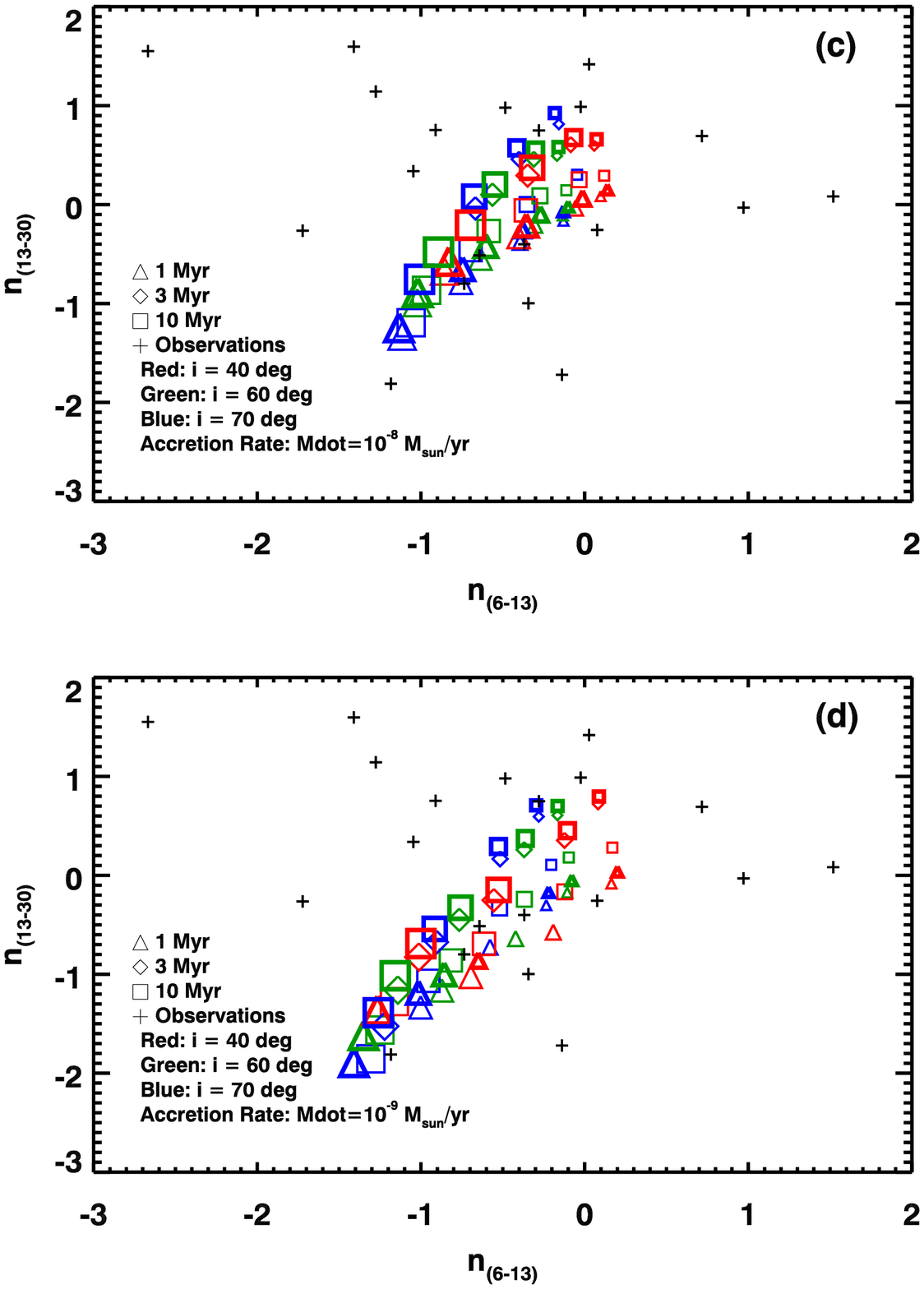}\\
{Fig. 10. --- Continued.}
\end{center}

\clearpage

%
\begin{figure}
\includegraphics[scale=0.8,angle=90]{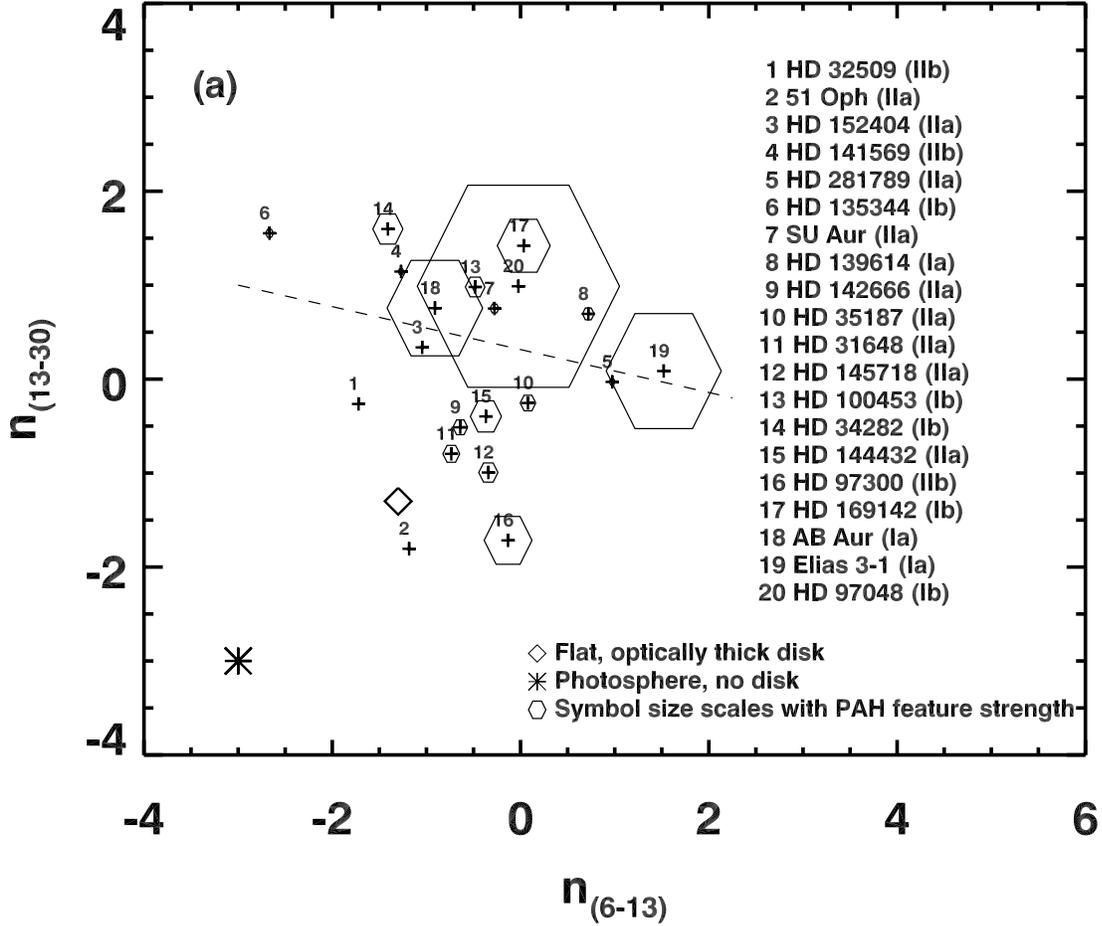} 
\caption{Continuum spectral indices of our sample stars. The dashed line indicates 
a boundary in our sample, below which the disks should have a more flattened geometry 
according to their Meeus et al. (2001) Group II classifications. Above the dashed line the disks 
are mostly Group I (with the exceptions of HD 141569 and SU Aur), which should have more
radially flared disks. All else being equal, increasing disk inclination should move sources towards the right in this diagram. Hydrocarbon ring (hexagon) symbol sizes scale with: (a) the {\it total} (integrated) 5-13~\mum\ PAH luminosity of the sources normalized to a distance of 100 pc from the observer; (b) the 6.2~\mum\ PAH feature luminosity {\it alone} normalized to a source distance of 100 pc from the observer; (c) the total PAH feature luminosity normalized to $L_{star}$, i.e symbol size is proportional to $[\Sigma F_{PAH}]/L_{star}$; and (d) $L_{star}$ only.}
\end{figure}
\clearpage
\begin{center}
\includegraphics[scale=0.8,angle=90]{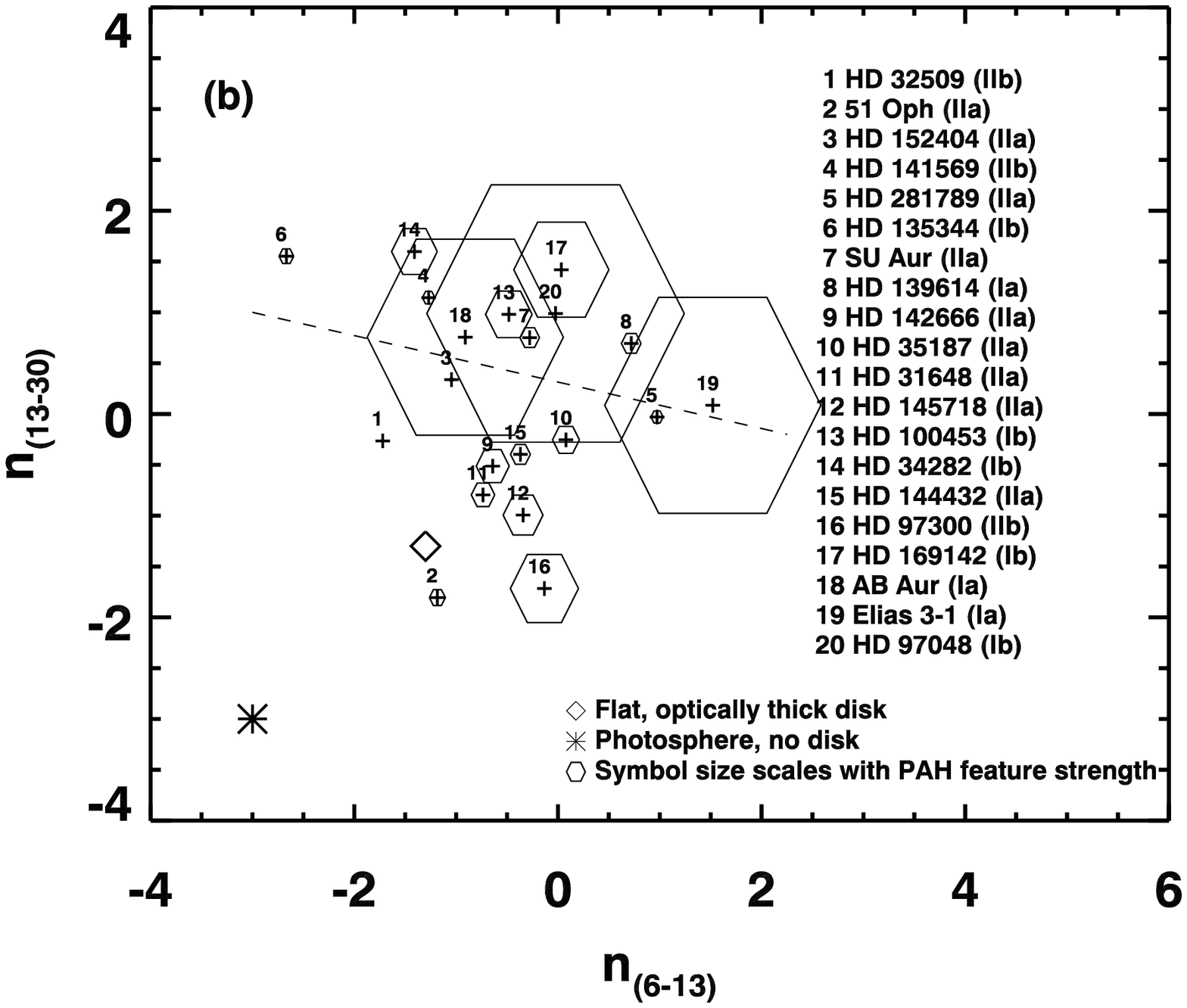}\\
{Fig. 11. --- Continued.}
\end{center}
\clearpage
\begin{center}
\includegraphics[scale=0.8,angle=90]{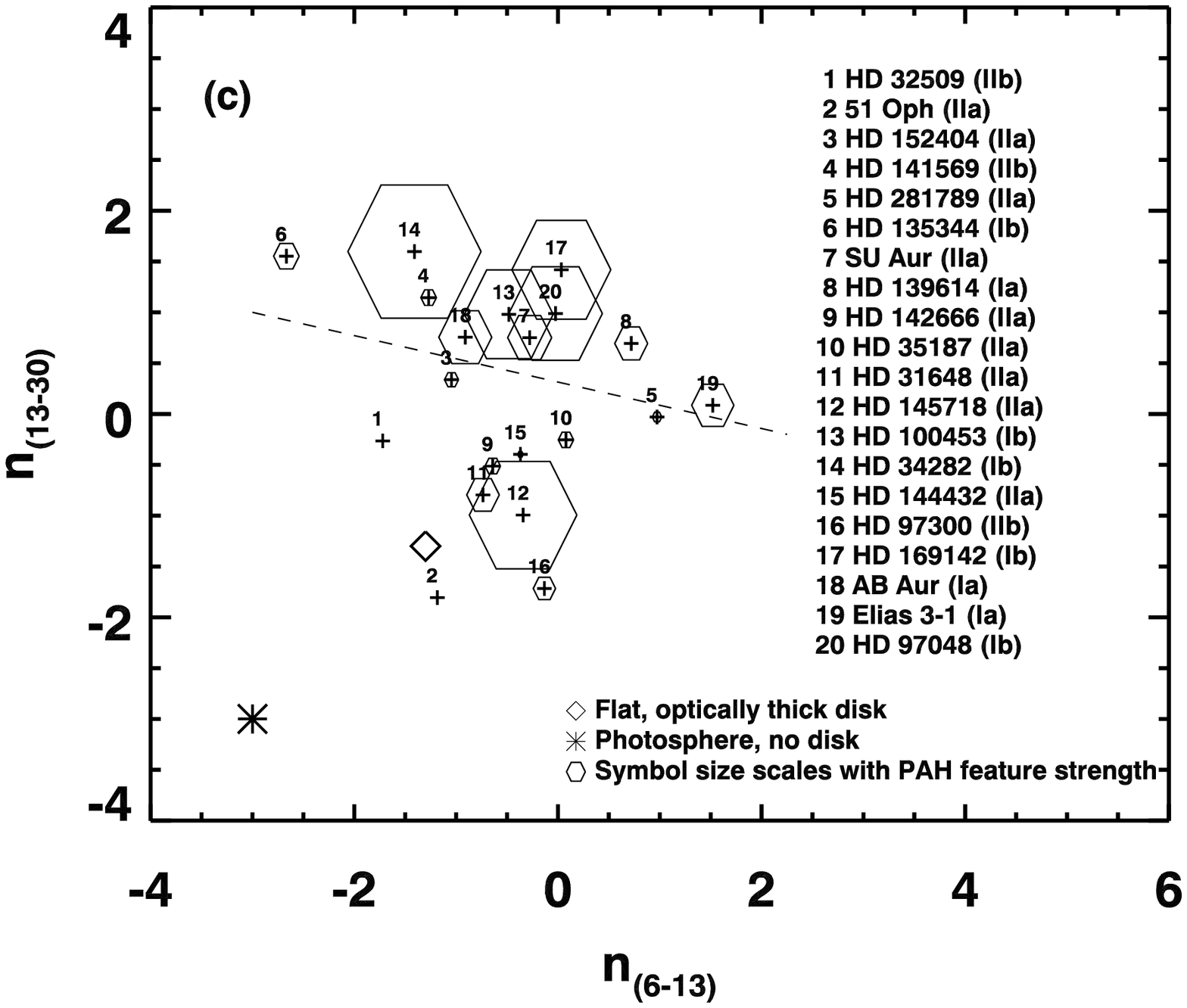}\\
{Fig. 11. --- Continued.}
\end{center}
\clearpage
\begin{center}
\includegraphics[scale=0.8,angle=90]{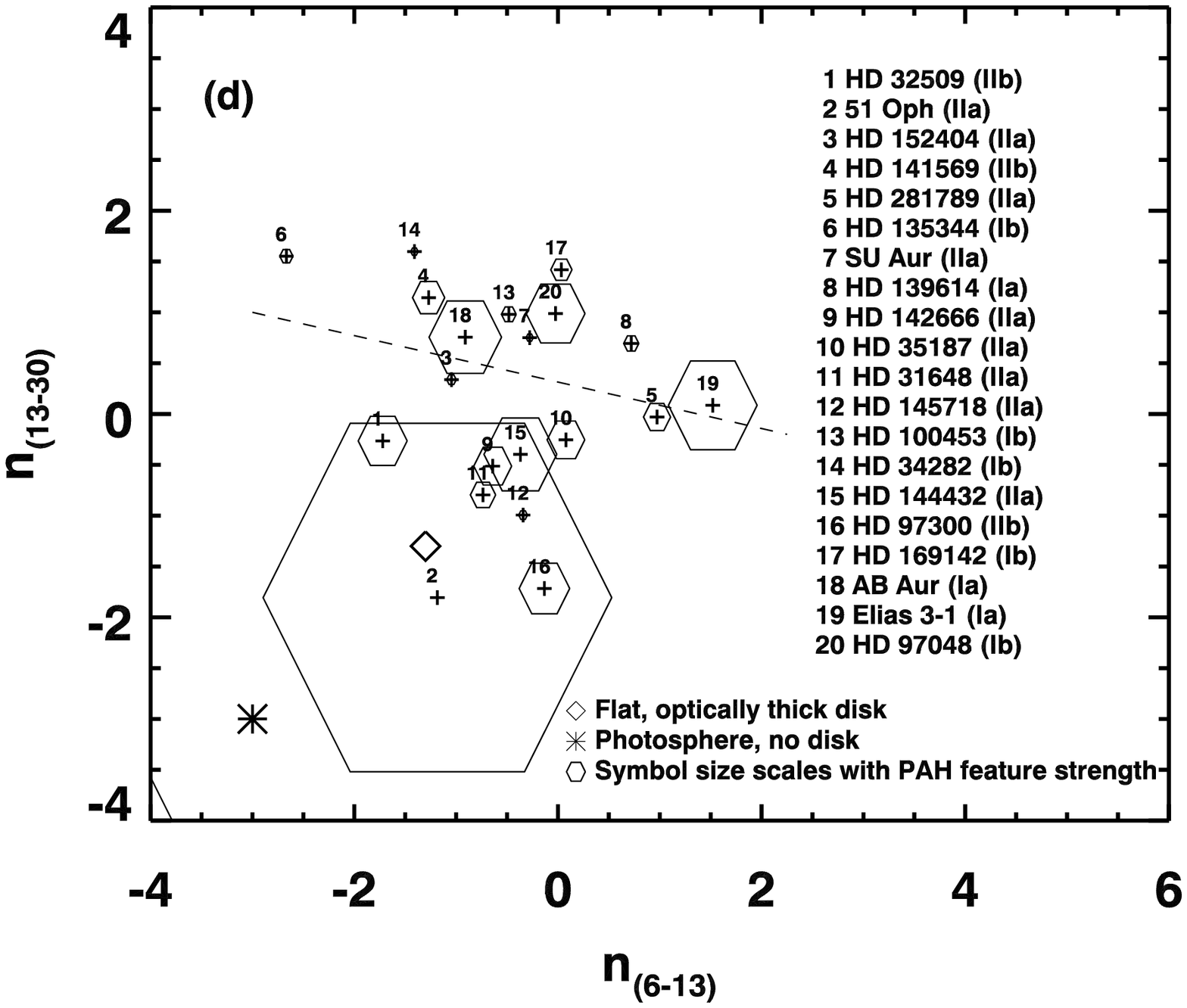}\\
{Fig. 11. --- Continued.}
\end{center}

\clearpage

%

\begin{figure}
\figurenum{12} 
\includegraphics[angle=0, scale=0.8]{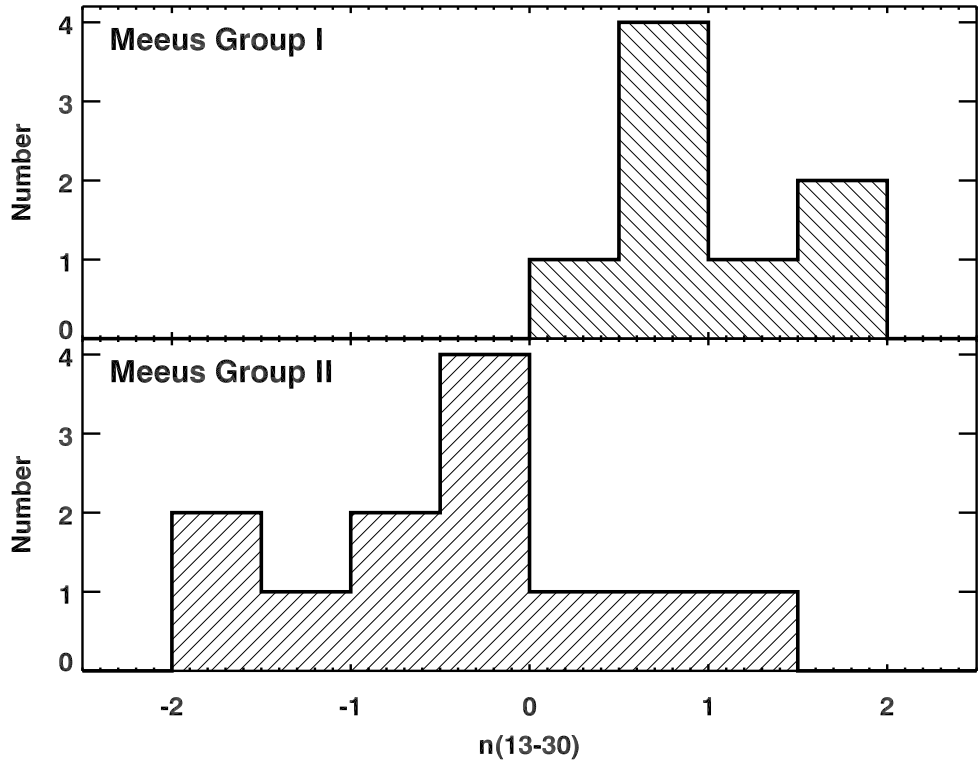}

\caption{Distribution of continuum color indices among Meeus Groups I \& II for our sample of HAeBe stars.} 

\end{figure}

\end{document}